\documentclass[12pt, draftclsnofoot, onecolumn]{IEEEtran}
\usepackage{amsmath,graphics,amssymb,epsfig,subfigure,rotating}
\usepackage{epic,eepic,latexsym}
\usepackage{algorithm}
\usepackage{algorithmic}
\usepackage{float}
\usepackage{multirow}
\usepackage{array}
\usepackage{cite}
\usepackage{setspace}

\newlength{\capwidth}
\newtheorem{Theorem}{Theorem}
\newtheorem{Corollary}{Corollary}
\newtheorem{Definition}{Definition}
\newtheorem{Lemma}{Lemma}

\newtheorem{Property}{Property}
\newtheorem{Proposition}{Proposition}

\newcommand{\yv}{{\mathbf y}}

\newcommand{\Yv}{{\mathbf Y}}
\newcommand{\Hv}{{\mathbf H}}
\newcommand{\Dv}{{\mathbf D}}
\newcommand{\Av}{{\mathbf A}}
\newcommand{\xv}{{\mathbf x}}
\newcommand{\sv}{{\mathbf s}}
\newcommand{\Sv}{{\mathbf S}}
\newcommand{\Xv}{{\mathbf X}}

\newcommand{\Cv}{{\mathbf C}}
\newcommand{\Nv}{{\mathbf N}}

\newcommand{\hv}{{\mathbf h}}

\newcommand{\Bv}{{\mathbf B}}
\newcommand{\Qv}{{\mathbf Q}}
\newcommand{\wv}{{\mathbf w}}

\newcommand{\Iv}{{\mathbf I}}

\newcommand{\Gv}{{\mathbf G}}

\newcommand{\Uv}{{\mathbf U}}
\newcommand{\Vv}{{\mathbf V}}
\newcommand{\Zv}{{\mathbf Z}}

\newcommand{\be}{\begin{equation}}
\newcommand{\ee}{\end{equation}}
\newcommand{\bea}{\begin{eqnarray}}
\newcommand{\eea}{\end{eqnarray}}
\newcommand{\bdp}{\begin{displaymath}}
\newcommand{\edp}{\end{displaymath}}

\begin{document}
\title{\huge{Leakage Rate Analysis for Artificial Noise Assisted Massive MIMO with Non-coherent Passive Eavesdropper in Block-fading}}

\author{\IEEEauthorblockN{Changick Song, \emph{Member, IEEE}} \\
\thanks{
C. Song is with the Electronic Engineering, Korea National University of Transportation, Chungju, Korea, 27469 (e-mail: c.song@ut.ac.kr).
}
\vspace{-40pt}
} \maketitle


\begin{abstract}
 Massive MIMO is one of the salient techniques for achieving high spectral efficiency in next generation wireless networks.
 Recently, a combined strategy of the massive MIMO and the artificial noise (AN), namely, {\it AN assisted massive MIMO (ANAM)}
 has recently been actively investigated for security enhancement.
 However, most of previous studies on the ANAM have been built upon the full channel state information (CSI)
 assumption at the eavesdropper (ED),
 which may be too pessimistic to provide meaningful information on the security
 since the channel uncertainty of the ED may degrade its decoding ability.
 In this paper, we provide more sophisticated investigation on the performance of the ANAM system
 assuming that the CSI of the ED channels are unknown to both the BS and the ED or partially known to the ED.
 We measure the secrecy in terms of both the leakage rate to the ED and the secrecy rate to the legitimate users,
 and characterize their upper and lower bounds in the high SNR regime as a function of the number of ED antennas,
 the number of data and AN signal dimensions, and coherence time.
 Finally, from numerical results, we demonstrate the accuracy of our analysis and highlight the security potential of the ANAM system against
 the passive eavesdropping attack.
\end{abstract}
\vspace{-10pt}
\vspace{-0pt}
\section{Introduction}\label{sec:introduction}

Along with the growing prevalence of wireless radio technologies, the security has become a major social challenge for both personal and professional sphere.
Unlike the wired communications, however, the wireless security
is in general a challenging task, since the radio transmission has no physical boundary, and thus any receivers nearby can listen to the transmitted signals.
As the wireless link is also unreliable and severely constrained by energy and bandwidth, more sophisticated physical layer designs are required.
An important objective of the physical layer wireless security is to
protect such an radio interface by negligibly {\it low probability of interception (LPI)}
without relying on (but can be integrated with) the upper-layer crypto system.

Several classical solutions have already existed to achieve the LPI.
For example, a waveform can be designed at the base-station (BS) using furtive frequency/time hopped or spread spectrum signals \cite{CLing:13}.
However, it may not be secure enough in wideband systems due to reduced space for the spreading gain.
If the channel state information (CSI) of the eavesdropper (ED) is allowed at the BS,
we may apply the directional beamforming scheme so that the data signals do not appear at the ED.
However, it is difficult to achieve in practice because the EDs are typically passive so as to hide their existence.

To address such issues, there are several of LPI designs that do not rely on both the ED's CSI at the BS.
One is to broadcast artificial noise (AN) signals isotropically
within the null space of the legitimate receiver (LU) channels \cite{SGoel:08,AKhisti:10,PHLin:13,HMWang:15a,TYLiu:17}.
Another way is to utilize a large excess of BS antennas as in \cite{DKapetanovic:15},
in which the channel hardening effect automatically arises rendering stable and predictable channel condition to the LUs,
while nearly nullified channel gains to the ED.
In addition, a combined strategy of the massive MIMO and the AN, namely, {\it AN assisted massive MIMO (ANAM)} has recently been actively investigated
for security enhancement \cite{JZhu:14} \cite{JZhu:16}.

As for the CSI acquisition in the massive MIMO systems,
the time division duplexing (TDD) mode, in which the channel reciprocity is exploited to allow the BS to obtain CSI
through uplink channel training, is in general a better choice than the frequency division duplexing (FDD) mode.
This is because the downlink training in FDD yields prohibitively large overhead as the number of BS antennas increases \cite{ZJiang:15}.

In fact, no downlink training yields an advantage in terms of the security, because the coherent detection is not allowed at the ED.
Unlike the LUs that can enjoy the deterministic channel condition via the beamforming from the BS,
the best strategy that can be taken by the ED to intercept the messages is the blind estimation (or detection) schemes \cite{MFrikel:00} \cite{YChi:10}
whose performance is heavily dependent on the coherence time and dimensionality of the signal space.
Thus, it is expected that as the coherence time of the ED channel reduces or the number of BS antennas increases,
the security will be enhanced.
Nevertheless, most of current research on the ANAM has been built upon the full CSI assumption at the ED,
which gives rises to more pessimistic results than it actually is, and thus could be misleading.

In this paper, we provide more sophisticated investigation on the performance of the TDD-based ANAM systems
with the non-coherent passive ED in block fading, where the CSI of the ED channels are unknown to both the BS and the ED.
The goal of the paper is to find an analytic expression of the leakage rate that measures the amount of information leaked to the non-coherent ED.
Note that once the leakage rate is identified, the secrecy rate can be easily computed, which represents the maximum secure data rate to the LUs
with the aid of the wiretap codes \cite{MBloch:11} \cite{CLing:14}.
Similar security concepts of no downlink training have been developed for the artificial fast fading (AFF) schemes in \cite{XLi:07,YKozai:15,HMWang:15,CSong:18TVT},
but the analysis has been based on the full CSI assumption at the ED.
The contribution of the paper can be summarized as follows.

\begin{itemize}
  \item First, in Section \ref{sec:Non-coherent Leakage Rate Analysis}, we analyze the leakage rate to the non-coherent ED, namely, {\it `non-coherent leakage rate'} based on the high signal-to-noise ratio (SNR) approximation. Our analysis not only unveils the degrees of freedom (DoF) of the leakage rate, but also identifies upper and lower bounds of the constant terms that are irrespective of ED's SNR. The bounds become tight as the coherence time increases over the number of BS antennas. Interestingly, we observe that when we scale up the dimensionality of the signal and AN space to the level of coherence time, the leakage rate DoF converges to zero regardless of the number of ED antennas and even without the aid of the wiretap codes. Obviously, this is not the case for the coherent ED \cite{AKhisti:10}. Note that our analysis is different from the previous works for non-coherent MIMO capacity in \cite{TMarzetta:99,LZheng:02,SSamai:02} because the artificial noise has not been taken into account, which makes the problem more challenging.
  \item Next, in Section \ref{sec:coherent leakage rate with partial CSI}, we examine {\it `partially coherent leakage rate'} considering a situation where the BS uses a downlink training over a few of beamforming vectors like the {\it beamformed CSI-RS} in LTE-A \cite{HJi:16}. The situation may occur when the CSIs at the BS is imperfect or outdated so that the LUs need to estimate their effective channels more accurately. In this case, the ED may also overhear the training signals and perform the coherent detection. Nevertheless, the channel uncertainty still remains since the AN channels are still unknown. The leakage rate is investigated in terms of both a universal upperbound that holds for all SNRs, and the tight upper and lower bounds that hold at high SNR. It turns out that the universal bound is tight in the low SNR regime regardless of the coherence time, while a relatively long coherence time is needed for the tightness of the high SNR bounds.
  \item In Section \ref{sec:Discussion}, we provide useful insights into the design through in-depth discussion on the analytical performance of the ANAM systems. Specifically, we examine the required number of BS antennas for the zero-DoF leakage rate according to the coherence time of the ED channels, the artificial fast fading design that deliberately shortens the ED's channel coherence time, and the achievable secrecy-rate to the LUs considering the wiretap codes at the BS. Finally, in \ref{sec:Simulation Results}, we present extensive simulation results to confirm the accuracy of our analysis and demonstrate the robustness of the ANAM system to the passive eavesdropping attack.
\end{itemize}

{\it \textbf{Notations}}: Normal letters represent scalar quantities, boldface letters indicate vectors and boldface uppercase letters
designate matrices.
The superscripts $(\cdot)^\mathsf{T}$, $(\cdot)^*$, and $(\cdot)^\mathsf{H}$ stand for the transpose, conjugate, and conjugate transpose operations, and
$\mathbb{C}$ and $\mathbb{R}_+$ denote sets of complex and real positive numbers, respectively.
We denote by $\Iv_N$ an $N\times N$ identity matrix and by $\mathsf{E}[\cdot]$ the expectation.
We write $\textrm{Tr}\left(\Av\right)$ and $\det(\Av)$ for the trace and determinant of a matrix $\Av$, respectively.
We use $(\cdot)^+$ and $\mathcal{I}(x;y)$ to denote $\max(\cdot,0)$ and the mutual information between two random variables $x$ and $y$, respectively.
We write $h(\cdot)$ as differential entropy to the base $2$. The equivalence $\overset{d}{=}$ means the same distribution.

\section{System Model}\label{sec:System Model}


In this paper, we consider a multiuser downlink system where a BS having a large $M$ number of antennas supports $K$ LUs with a single antenna
in the presence of an ED with $N_E$ antennas.
Note that the multi-antenna ED can be interpreted as multiple cooperative EDs with a total of $N_E$ antennas.
Similarly, the proposed analytical results are generally applicable to the case of a single LU with $K$ antennas.
We adopt a Rayleigh flat fading model in which the baseband channels from the BS to the $k$-th LU and ED are respectively expressed by
$\hv_k^\mathsf{H}\in\mathbb{C}^{1\times M}$ and $\Gv\in\mathbb{C}^{N_E\times M}$ whose propagation coefficients
are {\it independent and identically distributed} (i.i.d.) complex Gaussian with zero mean and unit variance, i.e., $\mathcal{CN}(0,1)$.
We consider the block-fading where all channel coefficients remain constant for $T$ symbol periods, and change to new independent values in the next time period.

The BS operates in a TDD mode and has perfect knowledge of $\Hv=[\hv_1,\ldots,\hv_K]^\mathsf{H}$
utilizing the uplink training from the LUs and the channel reciprocity.
Meanwhile, the instantaneous CSIs of the ED channel $\Gv$ is completely unknown to the BS due to the passive eavesdropping of the ED,
although its statistical information may be known to the BS.\footnote{
Recent advances in physical layer security have shown that it is actually possible for the BS to estimate the geometrical or stochastic information of the ED channels, even if the ED operates in a passive mode via detecting the inevitable power leakage of the ED's local oscillator \cite{AMukherjee:12} or utilizing the torch aided methods in \cite{YChoi:15}.}
The BS basically transmits no downlink training in order to prevent the ED from coherent detection, which means that
the CSI is unavailable at both the LUs and the ED.\footnote{Note that the case of partially coherent ED will also be covered later in Section \ref{sec:coherent leakage rate with partial CSI}.}
As in the conventional massive MIMO, the system is designed such that $M$ is much greater than $K$.
In contrast, $N_E$ could be arbitrarily large to be compared with $M$, given the situation where several multi-antenna EDs can cooperate.
For simplicity, we consider homogeneous users, i.e., each LU experiences the same received signal power on average and ignore the uplink phase duration
in the coherence time $T$.

As we have assumed that $M\gg K$, the extra antenna dimension at the BS can be exploited for AN transmission to interfere with the EDs.
Specifically, during each coherence interval,
the BS generates and transmits an $M\times T$ dimensional signal matrix $\Xv\in\mathbb{C}^{M\times T}$ as
\bea\label{eq:transmit signal}
\Xv=\alpha\Hv^\dagger\Sv+\beta\Vv_{H_2,N_J}\Nv,
\eea
where $\Sv\triangleq[\sv_1,\ldots,\sv_K]^\mathsf{H}\in\mathbb{C}^{K\times T}$ denotes the message signals whose $k$-th row $\sv_k^\mathsf{H}\in\mathbb{C}^{1\times T}$
carries a message to the $k$-th LU, $\Hv^{\dagger}\triangleq\sqrt{M}\Hv^\mathsf{H}(\Hv\Hv^\mathsf{H})^{-1}$ refers to the pseudo-inverse of $\Hv$,
and $\Nv\in\mathbb{C}^{N_J\times T}$ contains the $N_J$ dimensional artificial jamming noise with $N_J\leq M-K$.
Here, $\Vv_{H_2,N_J}\in\mathbb{C}^{M\times N_J}$ represents an orthonormal matrix with $\Vv_{H_2,N_J}^H\Vv_{H_2,N_J}=\Iv_{N_J}$ and $\Hv\Vv_{H_2,N_J}=\mathbf{0}$, which
is attainable from the following singular value decomposition (SVD)
\bea\label{eq:SVD}
\Hv=\Uv_{H}[\begin{array}{cc}
      \mathbf{\Lambda}_{H} & \mathbf{0}
    \end{array}]
    \Vv_H^\mathsf{H},
\eea
where $\Uv_{H}\in\mathbb{C}^{K\times K}$ and $\Vv_H\triangleq[\Vv_{H_1}~\Vv_{H_2}]\in\mathbb{C}^{M\times M}$ denote unitary matrices
with $\Vv_{H_1}\in\mathbb{C}^{M\times K}$ and $\Vv_{H_2}\in\mathbb{C}^{M\times (M-K)}$, and
$\mathbf{\Lambda}_{H}\in\mathbb{C}^{K\times K}$ refers to a diagonal matrix having non-zero singular values.
Then, $\Vv_{H_2,N_J}$ in (\ref{eq:transmit signal}) is defined by the first $N_J$ columns of $\Vv_{H_2}$.
It is seen that in each symbol time, the BS can transmit at most $M-K$ dimensional AN signals isotropically in the null subspace of the main channel $\Hv$.

It is assumed that $\Sv$ and $\Nv$ have i.i.d. Gaussian entries $s_{ij}\sim\mathcal{CN}(0,1)$ and $n_{ij}\sim\mathcal{CN}(0,1)$, respectively
to maximize the information rate to the LUs while minimizing the leakage rate to the ED for a given $\Sv$.
The scaling factors $\alpha\in\mathbb{R}_+$ and $\beta\in\mathbb{R}_+$ distribute the power to the message and AN signals,
subject to a power constraint $\text{Tr}(\mathsf{E}[\Xv\Xv^\mathsf{H}])=MT$ where
\bea\label{eq:power normalization}
\text{Tr}\left(\mathsf{E}[\Xv\Xv^\mathsf{H}]\right)&=&TM\alpha^2\mathsf{E}\left[\text{Tr}\left((\Hv\Hv^\mathsf{H})^{-1}\right)\right]+\beta^2 N_JT,\nonumber\\
&=&\frac{MK}{M-K}\alpha^2+N_J\beta^2.
\eea
The last equality follows from $\mathsf{E}[(\Hv\Hv^\mathsf{H})^{-1}]=\frac{1}{M-K}\Iv_K$ \cite{DMaiwald:00}.
The values of $\alpha$ and $\beta$ are fixed over a codeword duration, and therefore assumed to be known to all nodes.

After the signal in (\ref{eq:transmit signal}) is transmitted, corresponding received signals at the $k$-th LU and ED can be respectively expressed by
\bea
\label{eq:received signal at LU}
\yv_k^\mathsf{H}&=&\alpha\sqrt{M}\sv_k^\mathsf{H}+\wv_k^\mathsf{H}\\
\label{eq:received signal at ED}
\text{and}~~\Yv_E&=&\Gv_1\Sv+\Gv_2\Nv+\Zv,
\eea
where $\wv_k^\mathsf{H}\in\mathbb{C}^{1\times T}$ and $\Zv\in\mathbb{C}^{N_E\times T}$
denote the Gaussian thermal noise having i.i.d entries $w_{kj}\sim\mathcal{CN}(0,\sigma_w^2)$ and $z_{ij}\sim\mathcal{CN}(0,\sigma_z^2)$, respectively.
Also, we define $\Gv_1\triangleq\alpha\Gv\Hv^{\dagger}\in\mathbb{C}^{N_E\times K}$ and
$\Gv_2\triangleq\beta\Gv\bar{\Vv}_{H,N_J}\in\mathbb{C}^{N_E\times N_J}$ as the effective channels of the ED.
We refer to $\text{SNR}_L=\frac{1}{\sigma_w^2}$ and $\text{SNR}_E=\frac{1}{\sigma_z^2}$ as the expected SNRs at each receive antenna
of the LUs and the ED, respectively.

In our system, the channel coding can be applied over multiple $n$ fading blocks.
We do not assume any wiretap codes until we discuss the secrecy rate.
The encoding function is thus a one-to-one mapping between the messages and the codewords.
Let us define $\mathcal{M}_k=\{1,\ldots,2^{nTR_k}\}$ as a message set for the $k$-th LU with a transmission rate $R_k$.
Then, the secrecy can be measured in terms of a mutual information $\mathcal{I}(\mathcal{M}_1,\ldots,\mathcal{M}_K;\Yv^{(n)}_E)=\mathcal{I}(\Sv^{(n)};\Yv^{(n)}_E)$ which measures the amount of information leaked to the ED.

Because of the independence among different coherence intervals, it is sufficient to study one coherence interval.
Therefore, equivalently we can measure the secrecy via leakage rate that is defined by
\bea\label{eq:leakage}
\mathcal{L}_{\text{noncoherent}}(K,N_E,N_J,T)=\frac{1}{T}\mathcal{I}(\Yv_E;\Sv).
\eea
In this paper, we call (\ref{eq:leakage}) {\it non-coherent leakage rate},
because neither the BS nor the ED knows the CSIs of both $\Gv_1$ and $\Gv_2$.

The exact values of the leakage information can be formulated as
\bea\label{eq:exact leakage}
I(\Sv;\Yv_E)&=&h(\Yv_E)-h(\Yv_E|\Sv)\nonumber\\
&=&\mathsf{E}\left[\log\frac{P(\Yv_E|\Sv)}{P(\Yv_E)}\right]\nonumber\\
&=&\int d\Sv P(\Sv)\int d\Yv_E P(\Yv_E|\Sv)\log\left\{\frac{P(\mathbf{Y}_E|\mathbf{S})}{\int d\mathbf{S}' P(\mathbf{S}')P(\mathbf{Y}_E|\mathbf{S}')}\right\}\nonumber\\
&=&\int d\Sv P(\Sv)\int d\Yv_E \int d\Nv'P(\Nv')P(\Yv_E|\Sv,\Nv')\nonumber\\
&&\times\log\left\{\frac{\int d\mathbf{N}' P(\mathbf{N}')P(\mathbf{Y}_E|\mathbf{S},\mathbf{N}')}{\int d\mathbf{S}' P(\mathbf{S}')\int d\mathbf{N}'P(\mathbf{N}')P(\mathbf{Y}_E|\mathbf{S}',\mathbf{N}')}\right\}.
\eea
A direct evaluation of the mutual information in (\ref{eq:exact leakage}) is prohibitive even if the numerical methods are utilized,
because the exact distribution of $P(\Yv_E|\Sv,\Nv)$ is unknown and a myriad of integrals over multi-dimensional complex space are computationally intractable whenever the number of components of the signal matrices $\Sv$ and $\Nv$ are much greater than one in the massive MIMO systems.
The goal of the paper is to find an analytic expression of the leakage-rate, which enables us to evaluate the leakage rate without computing (\ref{eq:exact leakage})
for many cases of interest in the ANAM systems.

\section{Preliminaries}\label{sec:Preliminary}
In this section, we introduce some useful results and approximations that will be used for our mathematical derivations throughout the paper.

\subsection{Ergodic leakage rate}
Supposing that the effective ED channels $\Gv_1$ and $\Gv_2$ are perfectly known to the ED (but not at the BS),
the ergodic secrecy capacity was investigated in \cite{AKhisti:10a}.
In the following lemma, we reinterpret some of the results in terms of the leakage rate considering the i.i.d. Gaussian input and AN signals.

\begin{Lemma}[Ergodic Leakage Rate]\label{Lemma:Lemma1}
  For the given knowledge on $\bar{\Gv}\triangleq[\Gv_1~\Gv_2]\in\mathbb{C}^{N_E\times \bar{M}}$ with $\bar{M}=N_J+K$,
  $\Yv_E$ is Gaussian. Thus, the ergodic leakage rate can be computed and approximated at high $\text{SNR}_E$ as
  \bea\label{eq:ergodic MI}
  \mathcal{L}_{\text{ergodic}}(K,N_E,N_J)&\triangleq&\frac{1}{T}\mathcal{I}(\Yv_E;\Sv|\bar{\Gv})\nonumber\\
  &=&\mathsf{E}\Big[\log\det\left(\bar{\Gv}\bar{\Gv}^H+\sigma_z^2\Iv_{N_E}\right) -\log\det\left(\Gv_2\Gv_2^H+\sigma_z^2\Iv_{N_E}\right)\Big]\\
  &\overset{\sigma_z^2\rightarrow0}{=}&\min((N_E-N_J)^+,K)\log\text{SNR}_E+c_{K,N_E,N_J,\text{ergodic}}+o(1)\nonumber
  \eea
  where $o(1)$ represents the vanishing terms as $\sigma_z^2\rightarrow0$ and
  $c_{K,N_E,N_J,\text{ergodic}}$ denote a constant value irrespective of $\sigma_z^2$, which is defined by
  \bea
  c_{K,N_E,N_J,\text{ergodic}}&=&\mathsf{E}
  \bigg[\sum_{i=1}^{\min(\bar{M},N_E)}\log\lambda_{\bar{G},i}^2-\sum_{i=1}^{\min(N_E,N_J)}\log\lambda_{G_2,i}^2\bigg]\nonumber
  \eea
  with $\lambda_{\bar{G},i}$ for $i=1,\ldots,\min(\bar{M},N_E)$ and $\lambda_{G_2,j}$ for $j=1,\ldots,\min(N_E,N_J)$ being the non-zero singular values of $\bar{\Gv}$ and $\Gv_2$, respectively.
\end{Lemma}
\begin{IEEEproof}
By applying the SVD to $\bar{\Gv}$ and $\Gv_2$ in (\ref{eq:ergodic MI}), we have
\bea
  \mathcal{I}(\Yv_E;\Sv|\bar{\Gv})&=&T\mathsf{E}\left[\sum_{i=1}^{N_E}\log(\lambda_{\bar{G},i}^2+\sigma_z^2)-\sum_{i=1}^{N_E}\log(\lambda_{G_2,i}^2+\sigma_z^2)\right]\nonumber\\
  &=&-T\min((N_E-N_J)^+,K)\log\sigma_z^2\nonumber\\
  &&+T\mathsf{E}\left[\sum_{i=1}^{\min(\bar{M},N_E)}\log(\lambda_{\bar{G},i}^2+\sigma_z^2)-\sum_{i=1}^{\min(N_E,N_J)}\log(\lambda_{G_2,i}^2+\sigma_z^2)\right]\nonumber
  \eea
Finally, we obtain the lemma for $\sigma_z^2\rightarrow0$.
\end{IEEEproof}

We first find from Lemma \ref{Lemma:Lemma1} that the ergodic leakage rate has a DoF $\min((N_E-N_J)^+,K)$,
which means that the amount of information leaked to the coherent ED increases by $\min((N_E-N_J)^+,K)$ bps/Hz for each $3$dB $\text{SNR}_E$ increase.
If $N_E\leq N_J$, the DoF converges to zero, since all spatial dimension of the ED is corrupted by the AN.
In this case, thus, the leakage rate will be saturated at high $\text{SNR}$.
In contrast, when the ED has a sufficiently large number of antennas such that $N_E>\bar{M}$, the DoF increases up to $K$.
Therefore, when the CSI is allowed at the ED, a number of ED antennas could be a serious security threat despite the assistance of the AN.

\subsection{Large system approximation}

In this subsection, we examine the distributions of effective ED channels $\Gv_1$ and $\Gv_2$ in (\ref{eq:received signal at ED}).
\begin{Property}\label{Property:Property1}
  The entries of $\Gv_2$ are i.i.d. $\mathcal{CN}(0,\beta^2)$ and independent of $\Gv_1$.
\end{Property}
\begin{IEEEproof}
The ED channel $\Gv$ is {\it isotropically distributed (i.d.)}, since its distribution is invariant under rotation.
Thus, for an $M\times M$ unitary matrix $\Vv_H=[\Vv_{H_1}~\Vv_{H_2}]$, the entries of $\beta\Gv\Vv_H$ are still i.i.d. $\mathcal{CN}(0,\beta^2)$,
which implies that $\Gv_2=\beta\Gv\Vv_{H_2,N_J}$ has also i.i.d. $\mathcal{CN}(0,\beta^2)$ entries and is independent of $\Gv\Vv_{H_1}$.
Now, by applying the SVD in (\ref{eq:SVD}) to $\Gv_1$, we have an equality $\Gv_1=\alpha\Gv\Vv_{H1}\mathbf{\Lambda}^{-1}_H\Uv_H^H$.
Since $\Gv_2$ is independent of $\Gv\Vv_{H1}$ (the first half of $\Gv_1$), the remaining problem is to show that $\Gv_2$ is also independent of $\mathbf{\Lambda}^{-1}_H\Uv_H^H$.
This can be easily proved because $\Gv$ is independent of both $\mathbf{\Lambda}_H$ and $\Uv_H$, and $\Vv_{H2,N_J}$ is i.d. in the Stiefel manifold for any given $\mathbf{\Lambda}_H$ and $\Uv_H$.
Thus, $\Gv_2$ is independent of $\Gv_1$.
\end{IEEEproof}

\begin{Property}\label{Property:Property2}
  The entries of $\Gv_1$ are approximated to i.i.d. $\mathcal{CN}(0,\alpha^2)$ as $\frac{M}{K}\rightarrow\infty$.
\end{Property}
\begin{IEEEproof}
By the law of large numbers, we have $\lim_{\frac{M}{K}\rightarrow\infty}\frac{1}{M}\mathbf{H}\mathbf{H}^\mathsf{H}=\Iv_{K}$,
which implies that $\frac{1}{\sqrt{M}}\mathbf{\Lambda}_{H}\overset{M/K\rightarrow\infty}{=}\Iv_{K}$
and $\Gv_1\overset{M/K\rightarrow\infty}{=}\alpha\Gv\Vv_{H_1}\Uv_{H}^{\mathsf{H}}\overset{d}{=}\alpha\Gv\Vv_{H_1}$, and the proof is concluded.
\end{IEEEproof}

\begin{figure}
\begin{center}
\includegraphics[width=4.3in]{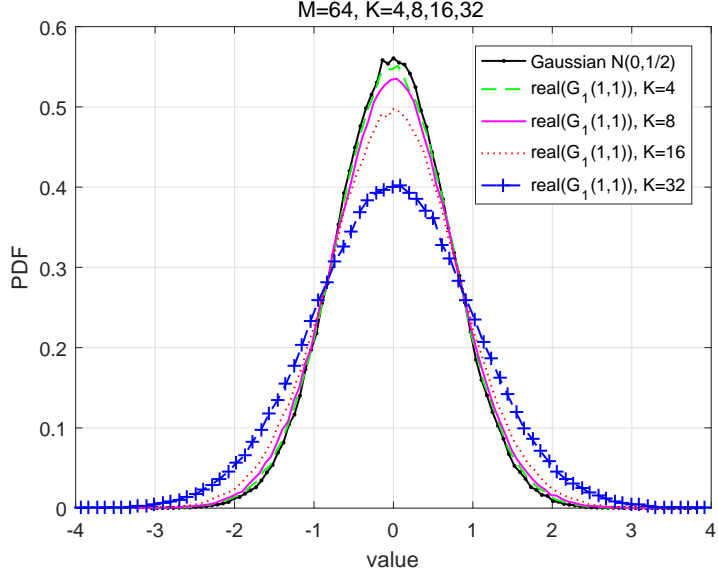}
\end{center}
\caption{Probability density function (PDF) for the $(1,1)$-th real component of $\Gv_1$ with $\alpha=1$. \label{Fig:figure0.eps}}
\end{figure}

Figure \ref{Fig:figure0.eps} demonstrates our statement in Property \ref{Property:Property2}.
We confirm from the figure that as $\frac{M}{K}$ increases, each element of $\Gv_1$ gets close to the corresponding Gaussian distribution.
It is observed that only a small value of $\frac{M}{K}$, e.g., $4$ or more, is sufficient for the approximation.

Now, from the large system approximation, i.e., $\frac{1}{M}\mathbf{H}\mathbf{H}^\mathsf{H}\overset{M/K\rightarrow\infty}{=}\Iv_{K}$,
the power constraint in (\ref{eq:power normalization}) asymptotically equal to
\bea\label{eq:power constraint approx}
\alpha^2K+\beta^2 N_J=M.
\eea
From now on, we consider (\ref{eq:power constraint approx}) when we determine the power distribution factors $\alpha$ and $\beta$.

\subsection{Mathematical definitions}
In what follows, we briefly introduce some mathematical definitions that will be used throughout the paper.

\begin{Definition}[Stiefel Manifold \cite{WMBoothby:86}]\label{Def:Definition1}
  The Stiefel manifold $S(T,M)$ for $T\geq M$ is defined as the set of all unitary $M\times T$ matrices
  $S(T,M)=\{\Uv\in\mathbb{C}^{M\times T}|\Uv\Uv^{\mathsf{H}}=\Iv_M\}$ and the total volume of the Stiefel manifold is computed as
  \bea
  |S(T,M)|=\prod_{i=T-M+1}^{T}\frac{2\pi^i}{(i-1)!}.\nonumber
  \eea
\end{Definition}

\begin{Definition}[Grassmann Manifold \cite{WMBoothby:86}]\label{Def:Definition2}
  The Grassmann manifold $G(T,M)$ for $T\geq M$ is defined as the quotient space of $S(T,M)$,
  which represents the set of all $M$-dimensional subspaces of $\mathbb{C}^T$.
  The volume of the Grassmann manifold equals
  \bea
  |G(T,M)|=\frac{|S(T,M)|}{|S(M,M)|}=\frac{\prod_{i=T-M+1}^{T}\frac{2\pi^i}{(i-1)!}}{\prod_{i=1}^{M}\frac{2\pi^i}{(i-1)!}}.\nonumber
  \eea
\end{Definition}

\begin{Definition}[Digamma Function \cite{PMLee:97}]\label{Def:Definition3}
  In mathematics, the digamma function $\varphi(x)\triangleq\frac{d}{dx}\ln\Gamma(x)$ is defined as the logarithmic derivative of the gamma function
  $\Gamma(x)\triangleq\int_{0}^{\infty}z^{x-1}e^{-z}dz$. Then, for a random Gaussian matrix $\Sv\in\mathbb{C}^{M\times T}$ whose entries are i.i.d. $\mathcal{CN}(0,1)$ for $T\geq M$,
  the following equivalence holds:
  \bea\label{eq:expected logdet}
  \mathsf{E}[\log\det(\Sv\Sv^\mathsf{H})]=\log e\sum_{i=1}^{M}\mathsf{E}[\ln\Gamma_{T-i+1,1}]
  =\log e\sum_{i=1}^{M}\varphi(T-i+1),
  \eea
  where $\Gamma_{(k,\theta)}$ denotes the gamma random variable with shape and scale parameters $k$ and $\theta$, respectively, and the last equality is due to the equivalence $\mathsf{E}[\ln\Gamma_{k,\theta}]=\varphi(k)-\ln\theta$. For an integer $x$, $\varphi(x)$ can be computed as $\varphi(x)=-\gamma+\sum_{p=1}^{x-1}\frac{1}{p}$ where $\gamma\simeq0.57721566$ is Euler's constant.
  The result in (\ref{eq:expected logdet}) will be periodically utilized throughout the paper.
\end{Definition}

\section{Non-coherent Leakage Rate Analysis}\label{sec:Non-coherent Leakage Rate Analysis}

In this section, we investigate high SNR approximate to the non-coherent leakage rate in (\ref{eq:leakage}).
Throughout the section, we assume that $T\geq\bar{M}$ and the ED channels are unknown to both the ED and the BS.
First, to compute $h(\Yv_E)$, we observe that $\Yv_E$ in (\ref{eq:received signal at ED}) can be rephrased by
\bea\label{eq:received signal at ED ver2}
\Yv_E=\bar{\Gv}\bar{\Xv}+\Zv
\eea
where $\bar{\Xv}\triangleq[\Sv^\mathsf{T}~\Nv^\mathsf{T}]^\mathsf{T}$ corresponds to a virtual input signal
having i.i.d. $\mathcal{CN}(0,1)$ entries and the effective ED channels $\bar{\Gv}=[\Gv_1~\Gv_2]$ is unknown to anybody.
Then, $h(\Yv_E)$ can be computed as in the following theorem.

\begin{Proposition}\label{Proposition:Proposition1}
  For $T\geq\bar{M}$, $h(\Yv_E)$ can be computed at high $\text{SNR}_E$ ($\sigma_z^2\rightarrow0$) as
  \bea\label{eq:theorem1}
  h(\Yv_E)=h(\hat{\Gv}\mathbf{\Lambda}_{\bar{X}}\Qv)+(T-\bar{M})\mathsf{E}\left[\sum_{i=1}^{\min(\bar{M},N_E)}\log\lambda_{\bar{G}\bar{X},i}^2\right]
  +\log\frac{\big|G\big(T,\min(\bar{M},N_E)\big)\big|}{\big|G\big(\max(\bar{M},N_E),N_E\big)\big|}\nonumber\\
  +(N_E-\bar{M})^+(T-\bar{M})\log\pi e\sigma_z^2+o(1),~~~
  \eea
  where $\hat{\Gv}\triangleq\bar{\Gv}\Uv_{\bar{X}}$ with $\Uv_{\bar{X}}\in\mathbb{C}^{\bar{M}\times\bar{M}}$ and $\mathbf{\Lambda}_{\bar{X}}\in\mathbb{C}^{\bar{M}\times\bar{M}}$
  denoting a unitary matrix and a diagonal matrix that stem from the SVD of
  $\bar{\Xv}=\Uv_{\bar{X}}\mathbf{\Lambda}_{\bar{X}}\Vv_{\bar{X}}^\mathsf{H}$, respectively,
  $\Qv\in\mathbb{C}^{\bar{M}\times\bar{M}}$ is an i.d. random unitary matrix that is independent of $\bar{\Gv}\bar{\Xv}$,
  and $\lambda_{\bar{G}\bar{X},i}$ is the $i$-th non-zero singular value of $\bar{\Gv}\bar{\Xv}$.
\end{Proposition}
\begin{IEEEproof}
See Appendix \ref{Apx:Appendix A}.
\end{IEEEproof}

In fact, the study on the differential entropy of $\Yv_E$ in (\ref{eq:received signal at ED ver2}) is not new \cite{TMarzetta:99} \cite{LZheng:02}.
Proposition \ref{Proposition:Proposition1} is however tailored for our purposes,
because the input distribution is fixed to i.i.d. complex Gaussian rather than the unitary matrix as is the case for non-secure communication systems.
In this proposition, we are also able to find a single compact expression of the non-coherent differential entropy, which is applicable to all antenna configurations with $T\geq\bar{M}$.

\begin{Lemma}\label{Lemma:Lemma2}
  When $\bar{\Gv}$ follows Properties \ref{Property:Property1} and \ref{Property:Property2},
  $h(\hat{\Gv}\mathbf{\Lambda}_{\bar{X}}\Qv)$ in (\ref{eq:theorem1}) is bounded by
  \bea\label{eq:lemma3}
     N_E(K\log\pi e\alpha^2+N_J\log\pi e\beta^2)+N_E\sum_{i=1}^{\bar{M}}\varphi(T-i+1)\log e~~~~~~~~~~~\nonumber\\
     \leq h(\hat{\Gv}\mathbf{\Lambda}_{\bar{X}}\Qv)\leq \bar{M}N_E\log\pi e T,
  \eea
  where $\varphi(\cdot)$ denotes the digamma function.
\end{Lemma}
\begin{IEEEproof}
See Appendix \ref{Apx:Appendix B}.
\end{IEEEproof}

Combining the results in Proposition \ref{Proposition:Proposition1} and Lemma \ref{Lemma:Lemma2},
we can establish the upper and lower bounds of $h(\Yv_E)$.
For example, provided that $K+N_J=M$ and $\alpha^2=\beta^2=1$,
the gap between the bounds in (\ref{eq:lemma3}) reduces to
\bea\label{eq:entropy gap}
MN_E\log T-N_E\sum_{i=1}^{M}\varphi(T-i+1)\log e~~~~~~~~~~~~~~~~~~~~\nonumber\\
\overset{T\rightarrow\infty}{=} N_E\left(M\log T-\sum_{i=1}^{M}\log(T-i+1)\right),
\eea
where the approximation holds due to the fact that
$\lim_{a\rightarrow\infty}\varphi(a)=\ln a$.
From (\ref{eq:entropy gap}), the gap will become smaller as the coherence time $T$ increases.
Therefore, the bounds in (\ref{eq:lemma3}) are asymptotically tight for $T\gg M$.
Note that the tightness is always preserved in terms of the DoF in (\ref{eq:theorem1}) regardless of $T$.

Now, to compute $h(\Yv_E|\Sv)$, let us rewrite $\Yv_E$ in (\ref{eq:received signal at ED}) as
\bea
\Yv_E^\mathsf{T}=\Nv^\mathsf{T}\Gv_2^\mathsf{T}+\Sv^\mathsf{T}\Gv_1^\mathsf{T}+\Zv^\mathsf{T}.\nonumber
\eea
Then, the system can be interpreted as the fictitious case of the BS sending a virtual signal matrix $\Gv_2^\mathsf{T}$ through a random propagation matrix $\Nv^\mathsf{T}$
with effective noise $\Sv^\mathsf{T}\Gv_1^\mathsf{T}+\Zv^\mathsf{T}$.
For a given $\Sv$, each column of the effective noise has an equal covariance matrix $\alpha^2\Sv^\mathsf{T}\Sv^*+\sigma_z^2\Iv_T$.
Thus, we can perform the noise whitening as
\bea\label{eq:whitening}
\bar{\Yv}_E^\mathsf{T}&\triangleq&\sigma_z(\alpha^2\Sv^\mathsf{T}\Sv^*+\sigma_z^2\Iv_T)^{-\frac{1}{2}}\Yv_E^\mathsf{T}\nonumber\\
&\overset{d}{=}&\sigma_z(\alpha^2\Sv^\mathsf{T}\Sv^*+\sigma_z^2\Iv_T)^{-\frac{1}{2}}\Nv^\mathsf{T}\Gv_2^\mathsf{T}+\Zv^\mathsf{T},
\eea
where the second equality holds true, because $\sigma_z(\alpha^2\Sv^\mathsf{T}\Sv^*+\sigma_z^2\Iv_T)^{-\frac{1}{2}}(\Sv^\mathsf{T}\Gv_1^\mathsf{T}+\Zv^\mathsf{T})\overset{d}{=}\Zv^\mathsf{T}$.

In the meantime, let us define the eigenvalue decomposition of $\Sv^\mathsf{T}\Sv^*$ as
\bea
\Sv^\mathsf{T}\Sv^*=\Vv_{S}
\bigg[
\begin{array}{cc}
  \mathbf{\Lambda}_S^2 & \mathbf{0} \\
  \mathbf{0} & \mathbf{0}
\end{array}
\bigg]\Vv_S^\mathsf{H}\nonumber
\eea
where $\mathbf{\Lambda}_S=\text{diag}\{\lambda_{S,1},\ldots,\lambda_{S,K}\}\in\mathbb{C}^{K\times K}$
represents a diagonal matrix containing non-zero singular values of $\Sv$ and
$\Vv_S\triangleq[\Vv_{S_1}~\Vv_{S_2}]\in\mathbb{C}^{T\times T}$ is a unitary matrix with
$\Vv_{S_1}\in\mathbb{C}^{T\times K}$ and $\Vv_{S_2}\in\mathbb{C}^{T\times(T-K)}$.
Now, applying the above decomposition to (\ref{eq:whitening}), it follows that
\bea\label{eq:received signal at ED ver3}
\bar{\Yv}_E^\mathsf{T}&\overset{d}{=}&\Vv_{S}
\bigg[
\begin{array}{cc}
  \sigma_z(\alpha\mathbf{\Lambda}_S)^{-1} & \mathbf{0} \\
  \mathbf{0} & \Iv_{T-K}
\end{array}
\bigg]\Vv_S^\mathsf{H}\Nv^\mathsf{T}\Gv_2^\mathsf{T}+\Zv^\mathsf{T}\nonumber\\
&=&\underbrace{\Vv_{S_2}\Vv_{S_2}^\mathsf{H}\Nv^\mathsf{T}\Gv_2^\mathsf{T}+\sigma_z\Vv_{S_1}(\alpha\mathbf{\Lambda}_S)^{-1}\Vv_{S_1}^\mathsf{H}\Nv^\mathsf{T}\Gv_2^\mathsf{T}}_{(a)}+\Zv^\mathsf{T}.
\eea

We observe that the resulting sum of $(a)$ has rank $N_J$, while the rank of $\Vv_{S_2}\Vv_{S_2}^\mathsf{H}\Nv^\mathsf{T}\Gv_2^\mathsf{T}$ equals $\min(T-K,N_J)$.
Therefore as long as $T-K\geq N_J$ (i.e., $T\geq\bar{M}$),
the second term of (\ref{eq:received signal at ED ver3}) becomes negligible as $\sigma_z\rightarrow0$.
In addition, as it is true that $\bar{\Yv}_E^\mathsf{T}\overset{d}{=}\Vv_S^\mathsf{H}\bar{\Yv}_E^\mathsf{T}$, we have
\bea
\bar{\Yv}_E^\mathsf{T}&\overset{d}{=}&
\bigg[
\begin{array}{c}
\Zv_1^\mathsf{T} \\
\Nv_2^\mathsf{T}\Gv_2^\mathsf{T}+\Zv_2^\mathsf{T}
\end{array}
\bigg],
\eea
where $\Zv_1=\Zv\Vv_{S_1}^\mathsf{*}\in\mathbb{C}^{N_E\times K}$ and
$\Zv_2=\Zv\Vv_{S_2}^\mathsf{*}\in\mathbb{C}^{N_E\times(T-K)}$ have i.i.d $\mathcal{CN}(0,\sigma_z^2)$ entries,
and the components of $\Nv_2=\Nv\Vv_{S_2}^\mathsf{*}\in\mathbb{C}^{N_J\times(T-K)}$ are i.i.d $\mathcal{CN}(0,1)$.

The differential entropy $h(\Yv_E|\Sv)=h(\Yv_E^\mathsf{T}|\Sv)$ is now computed by
\bea\label{eq:conditional entropy}
h(\Yv_E^\mathsf{T}|\Sv)&=&h(\bar{\Yv}_E^\mathsf{T}|\Sv)+N_E\mathsf{E}\left[\log\det\left(\alpha^2\Sv^\mathsf{T}\Sv^*+\sigma_z^2\Iv_T\right)\right]-N_ET\log\sigma_z^2\nonumber\\
&=&h(\Zv_1)+h(\Gv_2\Nv_2+\Zv_2)+N_E\mathsf{E}\left[\log\det\left(\alpha^2\Sv^\mathsf{T}\Sv^*+\sigma_z^2\Iv_T\right)\right]-N_ET\log\sigma_z^2\nonumber\\
&=&h(\Gv_2\Nv_2+\Zv_2)+ KN_E\log\pi e \sigma_z^2+N_E\mathsf{E}\left[\sum_{i=1}^K\log(\alpha^2\lambda_{S,i}^2)\right]\nonumber\\
&&+N_E(T-K)\log\sigma_z^2-N_ET\log\sigma_z^2\nonumber\\
&=&h(\Gv_2\Nv_2+\Zv_2)+N_E\mathsf{E}\left[\sum_{i=1}^K\log \lambda_{S,i}^2\right]+KN_E\log\pi e\alpha^2
\eea
where the first equality holds from the entropy scaling law $h(\Av\xv|\Av)=h(\xv)+\mathsf{E}[\log\det(\Av\Av^\mathsf{H})]$ for a square full-rank matrix $\Av$.
Finally, applying Proposition \ref{Proposition:Proposition1} to $h(\Gv_2\Nv_2+\Zv_2)$, we obtain $h(\Yv_E|\Sv)$ as summarized in Proposition \ref{Proposition:Proposition2}.

\begin{Proposition}\label{Proposition:Proposition2}
  Define $\Yv_E=\Gv_1\Sv+\Gv_2\Nv+\Zv$,
  where $\Gv_1\in\mathbb{C}^{N_E\times K}$, $\Gv_2\in\mathbb{C}^{N_E\times N_J}$, and $\Zv\in\mathbb{C}^{N_E\times T}$ have i.i.d. complex Gaussian entries with zero mean and variances $\alpha^2$, $\beta^2$, and $\sigma_z^2$, respectively,
  and the entries of $\Sv\in\mathbb{C}^{K\times T}$, $\Nv\in\mathbb{C}^{N_J\times T}$ are i.i.d $\mathcal{CN}(0,1)$.
  Also, we define $\Nv_2\in\mathbb{C}^{N_J\times (T-K)}$ as a matrix that consists of the last $T-K$ columns of $\Nv$.

  Then, for $T\geq\bar{M}$, the conditional entropy $h(\Yv_E|\Sv)$
  is computed at high $\text{SNR}_E$ ($\sigma_z^2\rightarrow0$) as
  \bea\label{eq:theorem2}
  h(\Yv_E|\Sv)=h(\hat{\Gv}_2\mathbf{\Lambda}_{N_2}\Qv)+(T-\bar{M})\mathsf{E}\left[\sum_{i=1}^{\min(N_J,N_E)}\log\lambda_{G_2N_2,i}^2\right]~~~~~~~~~~~~~~~\nonumber\\
  +N_E\sum_{i=1}^K\varphi(T-i+1)\log e+\log\frac{\big|G\big(T-K,\min(N_J,N_E)\big)\big|}{\big|G\big(\max(N_J,N_E),N_E\big)\big|}
  +KN_E\log\pi e\alpha^2\nonumber\\
  +(N_E-N_J)^+(T-\bar{M})\log\pi e\sigma_z^2+o(1),~~~~
  \eea
  where $\hat{\Gv}_2\triangleq\Gv_2\Uv_{N_2}$ with $\Uv_{N_2}\in\mathbb{C}^{N_J\times N_J}$ and $\mathbf{\Lambda}_{N_2}\in\mathbb{C}^{N_J\times N_J}$
  denoting a unitary matrix and a diagonal matrix that stem from the SVD of
  $\Nv_2=\Uv_{N_2}\mathbf{\Lambda}_{N_2}\Vv_{N_2}^\mathsf{H}$, respectively,
  $\Qv\in\mathbb{C}^{N_J\times N_J}$ is an i.d. random unitary matrix that is independent of $\Gv_2\Nv_2$,
  and $\lambda_{G_2N_2,i}$ is the $i$-th non-zero singular value of $\Gv_2\Nv_2$.
\end{Proposition}

Following the proof in Lemma \ref{Lemma:Lemma2}, $h(\hat{\Gv}_2\mathbf{\Lambda}_{N_2}\Qv)$ in (\ref{eq:theorem2}) is bounded by
  \bea\label{eq:lemma4}
     N_EN_J\log\pi e\beta^2+N_E\sum_{i=1}^{N_J}\varphi(T-K-i+1)\log e~~~~~~~~~\nonumber\\
     \leq h(\hat{\Gv}_2\mathbf{\Lambda}_{N_2}\Qv)\leq N_EN_J\log\pi e (T-K).
  \eea
Finally, by combining Propositions \ref{Proposition:Proposition1} and \ref{Proposition:Proposition2}, we can evaluate the high SNR approximate to the non-coherent leakage rate as summarized in the following theorem.

\begin{Theorem}[Non-coherent Leakage Rate]\label{Theorem:Theorem1}
  When the effective ED channels of both $\Gv_1$ and $\Gv_2$ are unknown to the ED with $T\geq\bar{M}$,
the non-coherent leakage rate is computed
  at high $\text{SNR}_E$ ($\sigma_z^2\rightarrow0$) as
  \bea\label{eq:proposition1}
     \mathcal{L}_{\text{non-coh}}(K,N_E,N_J,T)=\min((N_E-N_J)^+,K)\Big(1-\frac{\bar{M}}{T}\Big)\log\text{SNR}_E\nonumber\\
     +~c_{K,N_E,N_J,T,\text{non-coh}}+o(1),~~~~~~~~~~~~\nonumber
  \eea
  where $c_{K,N_E,N_J,T,\text{non-coh}}$ is defined and bounded by
  \bea\label{eq:Prop1 upper}
  c_{K,N_E,N_J,T,\text{non-coh}}&\triangleq&\frac{1}{T}h(\hat{\Gv}\mathbf{\Lambda}_{\bar{X}}\Qv)-\frac{1}{T}h(\hat{\Gv}_2\mathbf{\Lambda}_{N_2}\Qv)
  +d_{K,N_E,N_J,T,\text{non-coh}}\nonumber\\
  &\leq&\frac{N_E}{T}\Big\{K\log\pi e T+N_J\log\frac{T}{\beta^2}-\sum_{i=K+1}^{\bar{M}}\varphi(T-i+1)\log e\Big\}\nonumber\\
  &&+~d_{K,N_E,N_J,T,\text{non-coh}},\nonumber\\
  \label{eq:Prop1 lower}
  \text{and}~~c_{K,N_E,N_J,T,\text{non-coh}}&\geq&\frac{N_E}{T}\Big\{K\log\pi e\alpha^2+N_J\log\frac{\beta^2}{T-K}+\sum_{i=1}^{\bar{M}}\varphi(T-i+1)\log e\Big\}\nonumber\\
  &&+~d_{K,N_E,N_J,T,\text{non-coh}}~~~\nonumber
  \eea
  \bea
  \text{with}~~d_{K,N_E,N_J,T,\text{non-coh}}\triangleq  \Big(1-\frac{\bar{M}}{T}\Big)\mathsf{E}\bigg[\log\frac{\prod_{i=1}^{\min(\bar{M},N_E)}\lambda_{\bar{G}\bar{X},i}^2}{\prod_{i=1}^{\min(N_J,N_E)}\lambda_{G_2N_2,i}^2}\bigg]~~~~~~~~~~~~~~~~~~~~~~~~~~~~~~\nonumber\\
  -\frac{N_E}{T}\sum_{i=1}^K\varphi(T-i+1)\log e+\frac{1}{T}\log\bigg\{\frac{\big|G\big(T,\min(\bar{M},N_E)\big)\big|}{\big|G\big(\max(\bar{M},N_E),N_E\big)\big|}
  \frac{\big|G\big(\max(N_J,N_E),N_E\big)\big|}{\big|G\big(T-K,\min(N_J,N_E)\big)\big|}\bigg\}
\nonumber\\
  -\min((N_E-N_J)^+,K)\Big(1-\frac{\bar{M}}{T}\Big)\log\pi e-\frac{KN_E}{T}\log\pi e\alpha^2.\nonumber
\eea
\end{Theorem}
Note that the bounds (\ref{eq:Prop1 upper}) and (\ref{eq:Prop1 lower}) in Theorem \ref{Theorem:Theorem1} are based on the results in (\ref{eq:lemma3}) and (\ref{eq:lemma4}).

Focusing on the DoF, it is interesting to contrast the ergodic leakage rate in Lemma \ref{Lemma:Lemma1} and the non-coherent leakage rate in Theorem \ref{Theorem:Theorem1}.
As mentioned previously, the ergodic leakage rate is independent of the coherence time $T$,
because the ED can exploit its full spatial DoF by using the CSIs.
Therefore, a large number of ED antennas may be a security threat.
On the contrary, the result in Theorem \ref{Theorem:Theorem1} reveals that the channel uncertainty reduces the ED's spatial DoF by the factor of $(1-\frac{\bar{M}}{T})$.
This is because some of the space-time dimension of the ED channel should be consumed for channel uncertainty resolution.
Interestingly, if the BS increases the AN signal space dimension $N_J$ such that $\bar{M}=T$ or more,
we achieve a zero DoF in the leakage rate regardless of the number of ED antennas $N_E$.
Therefore, the proposed ANAM having a large number of transmit antennas with no downlink training is inherently robust to the passive eavesdropping
especially in a mobile environment where the coherence time $T$ is not too long.

Now, we investigate the constant value of the leakage rate in Theorem \ref{Theorem:Theorem1}.
Let us consider an example of $\bar{M}=M=T$ and $\alpha^2=\beta^2=1$.
In this case, the DoF equals zero, and thus the leakage rate upperbound is saturated to a constant value at high $\text{SNR}_E$ as
\bea\label{eq:saturated upperbound}
\mathcal{L}_{\text{non-coh}}(K,N_E,N_J,T)&\leq& N_E\log T-\frac{N_E}{T}\left\{\sum_{i=1}^{\bar{M}}\varphi(T-i+1)\log e\right\},\nonumber\\
&\leq&N_E\log e^\gamma T\nonumber
\eea
where $\gamma\simeq0.57721566$ denotes the Euler's constant.
We see that compared to the LUs' data rate at high SNR, the effect of a large antenna attack of the ED could be marginal even with respect to the constant values of the leakage rate.
The above bound is also independent of $K$, which means that
the increase of the number of LUs does not affect the amount of leakage to the ED.
An interesting point here is that we did not assume any wiretap codes or the CSI of the ED channels at the BS, both of which are practically difficult to realize.
Therefore, the ANAM is also of practical significance and has the potential of physical layer security.


\section{Partially Coherent Leakage Rate}\label{sec:coherent leakage rate with partial CSI}

In this section, we consider a practical situation where the BS transmits the downlink training precoded by $\Hv^\dagger$ to the LUs.
This is in fact useful when the knowledge of $\Hv$ at the BS is imperfect or outdated and therefore the LUs needs to estimate their effective channels more accurately.
However, the downlink training may also give the ED a chance of estimating its own effective channel $\Gv_1$.
Therefore, it is also practically of interest to investigate the partially coherent leakage rate supposing that the effective ED channel $\Gv_1$ is known to the ED
whereas the AN channel $\Gv_2$ is still unknown.

In order to estimate the fading coefficients of the $K$ different LUs, a training phase of duration should be no smaller than $K$ \cite{BHassibi:03}.
This represents the minimum cost for using the downlink training to the LUs, which reduces the effective coherence time for the data transmission from $T$ to $T^\prime\leq T-K$.
Thus, with the downlink training, the leakage rate can be formulated by
\bea\label{eq:leakage partialCSI}
\mathcal{L}_{\text{partial-coh}}(K,N_E,N_J,T^\prime)=\frac{1}{T^\prime}\mathcal{I}(\Yv_E^\prime;\Sv^\prime|\Gv_1)
\eea
where $\Yv_E^\prime=\Gv_1\Sv^\prime+\Gv_2\Nv^\prime+\Zv^\prime\in\mathbb{C}^{N_E\times T^\prime}$ denotes the received signals at the ED over the remaining coherence time $T^\prime$
with an input signal matrices $\Sv^\prime\in\mathbb{C}^{K\times T^\prime}$ (message signals) and $\Nv^\prime\in\mathbb{C}^{N_J\times T^\prime}$ (AN signals) both
having i.i.d. $\mathcal{CN}(0,1)$ entries, and an additive noise $\Zv^\prime\in\mathbb{C}^{N_E\times T^\prime}$ having i.i.d. $\mathcal{CN}(0,\sigma_z^2)$ entries.
As a worst case, we consider no channel estimation error at the ED.
Nevertheless, a direct evaluation of (\ref{eq:leakage partialCSI}) is difficult due to the hidden variables in $\Gv_2$.
To tackle the problem, we first investigate a universal upperbound that generally holds without any assumption.
Then, we obtain tighter upper and lower bounds under the assumption that $N_E\geq\bar{M}$, $T^\prime\gg N_J$, and $\sigma_z^2\rightarrow0$.

\subsection{Universal upperbound}

The following theorem evaluates the universal upperbound of the partially coherent leakage rate in (\ref{eq:leakage partialCSI}).

\begin{Theorem}\label{Theorem:Theorem2}
  Supposing that the ED perfectly knows $\Gv_1$ (not $\Gv_2$), the partially coherent leakage rate to the ED is universally upperbounded by
  \bea\label{eq:lemma2-00}
  \mathcal{L}_{\text{partial-coh}}(K,N_E,N_J,T^\prime)&\!\!\!\leq\!\!\!&\frac{1}{T^\prime}\mathcal{I}(\Yv_E^\prime;\Sv^\prime,\Gv_2,|\Gv_1,\Nv^\prime)
  -\frac{1}{T^\prime}\mathcal{I}(\Yv_E^\prime;\Gv_2|\Nv^\prime,\Gv_1,\Sv^\prime)\\
  \label{eq:lemma2-0}
  &\!\!\!=\!\!\!&\min(N_E,K)\left(1-\frac{N_J}{T^\prime}\right)^+\log\text{SNR}_E+c_{K,N_E,N_J,T^\prime,\text{partial-coh}}^{\text{univ}}
  \eea
  where $c_{K,N_E,N_J,T^\prime,\text{partial-coh}}^{\text{univ}}$ represents the terms that converges to a constant at high $\text{SNR}_E$ as
  \bea
  c_{K,N_E,N_J,T^\prime,\text{partial-coh}}^{\text{univ}}=\frac{1}{T^\prime}\mathsf{E}\bigg[\sum_{i=1}^{\min(T^\prime,N_J)}\sum_{j=1}^{\min(N_E,K)}
\log\bigg\{1+\frac{\lambda_{G_1,j}^2}{\beta^2\lambda_{N^\prime,i}^2+\sigma_z^2}\bigg\}\bigg]~~~~~~~~~~~\nonumber\\
+\left(1-\frac{N_J}{T^\prime}\right)^+\mathsf{E}\bigg[\sum_{j=1}^{\min(N_E,K)}\log(\lambda_{G_1,j}^2+\sigma_z^2)\bigg].\nonumber
  \eea
  with $\lambda_{N^\prime,i}$ for $i=1,\ldots,\min(T^\prime,N_J)$ and
  $\lambda_{G_1,j}$ for $j=1,\ldots,\min(N_E,K)$ denoting the non-zero singular values of $\Nv^\prime$ and $\Gv_1$, respectively,
  which is tight in the low $\text{SNR}_E$ region.
\end{Theorem}
\begin{IEEEproof}
See Appendix \ref{Apx:Appendix C}.
\end{IEEEproof}


\begin{Corollary}\label{corollary:corollary1}
At low $\text{SNR}_E$, the partially coherent leakage rate asymptotically equals
\bea\label{eq:interpret lemma2 2}
\mathcal{L}_{\text{partial-coh}}(K,N_E,N_J,T^\prime)&\overset{\sigma_z^2\rightarrow\infty}{=}\mathsf{E}\left[\log\left(\Iv_K+\text{SNR}_E\Gv_1^\mathsf{H}\Gv_1\right)\right],
\eea
\end{Corollary}
\begin{IEEEproof}
Assuming that $\sigma_z^2\rightarrow\infty$, we have
\bea
c_{K,N_E,N_J,T^\prime,\text{partial-coh}}^{\text{univ}}\overset{\sigma_z^2\rightarrow\infty}{=}\min(N_E,K)\Big(1-\frac{N_J}{T^\prime}\Big)^+\log\sigma_z^2~~~~~~~~~~~~~~~~~~~~~~~\nonumber\\
+\Big\{\frac{\min(T^\prime,N_J)}{T^\prime}+\Big(1-\frac{N_J}{T^\prime}\Big)^+\Big\}\mathsf{E}\Big[\sum_{j=1}^{\min(N_E,K)}
\log\Big(1+\frac{\lambda_{G_1,j}^2}{\sigma_z^2}\Big)\Big],
\eea
which leads to $\mathcal{L}_{\text{partial-coh}}(K,N_E,N_J,T^\prime)\overset{\sigma_z^2\rightarrow\infty}{\leq}\mathsf{E}\big[\sum_{j=1}^{\min(N_E,K)}\log\big(1+\frac{\lambda_{G_1,j}^2}{\sigma_z^2}\big)\big]$.
Since the universal bound in Theorem \ref{Theorem:Theorem2} is tight, we obtain the corollary.
\end{IEEEproof}

The results in Theorem \ref{Theorem:Theorem2} and Corollary \ref{corollary:corollary1} provide useful insight into the system.
First, observe that at low $\text{SNR}_E$, the partially coherent leakage rate converges to the ergodic capacity of the ED channel with no AN,
which implies that there may be no security gain that stems from the AN and its channel uncertainty.
In contrast, the universal bound in (\ref{eq:lemma2-0}) shows that  for the case of $\sigma_z^2<\infty$,
the AN and its channel uncertainty are still valid for emasculating the ED's decoding ability even if $\Gv_1$ is known to the ED,
since we can arbitrarily reduce (\ref{eq:lemma2-0}) by increasing $\beta^2$ or decreasing $T$.
Note that obviously the known CSI $\Gv_1$ is beneficial for the ED to intercept the messages compared to the non-coherent cases.
Thus, the bounds introduced in Theorem \ref{Theorem:Theorem2} can also be exploited as useful upperbound
for the non-coherent leakage rate $\mathcal{L}_{\text{noncoherent}}(K,N_E,N_J)$ in the previous section.

\subsection{Tight upper and lower bounds with high SNR approximation}\label{sec:Partial CSI at ED}

In this subsection, we investigate tighter upper and lower bounds for the partially coherent leakage rate based on the high SNR approximation.
By definition, we have
\bea\label{eq:leakage rate with partialCSI}
\mathcal{L}_{\text{partial-coh}}(K,N_E,N_J,T^\prime)&=&\frac{1}{T^\prime}h(\Yv_E^\prime|\Gv_1)-\frac{1}{T^\prime}h(\Yv_E^\prime|\Gv_1,\Sv^\prime)\nonumber\\
&=&\frac{1}{T^\prime}h(\Yv_E^\prime|\Gv_1)-\frac{1}{T^\prime}h(\Gv_2\Nv^\prime+\Zv^\prime)
\eea
Then, the high SNR approximate to the first and second terms of (\ref{eq:leakage rate with partialCSI})
can be attained following the arguments in Theorem \ref{Proposition:Proposition1} and \ref{Proposition:Proposition2}, respectively.
The result is summarized below. Detailed proof is omitted here for brevity.

\begin{Theorem}[Coherent Leakage Rate]\label{Theorem:Theorem3}
  Define $\Gv_{2,2}\in\mathbb{C}^{(N_E-K)\times N_J}$ as a random matrix having i.i.d $\mathcal{CN}(0,\beta^2)$ entries.
  Then, Supposing that $\Gv_1$ is known to the ED with $N_E\geq\bar{M}$, $T^\prime\geq N_J$, and $\sigma_z^2\rightarrow0$,
  the partially coherent leakage rate can be evaluated by
  \bea\label{eq:proposition3}
     \mathcal{L}_{\text{partial-coh}}(K,N_E,N_J,T^\prime)=K\left(1-\frac{N_J}{T^\prime}\right)^+\log\text{SNR}_E+c_{K,N_E,N_J,T^\prime,\text{partial-coh}}+o(1)\nonumber
  \eea
  where $c_{K,N_E,N_J,T^\prime,\text{partial-coh}}$ denotes the constant terms irrespective of $\text{SNR}_E$, which is defined and bounded by
  \bea
  c_{K,N_E,N_J,T^\prime,\text{partial-coh}}&\triangleq&\frac{1}{T^\prime}h(\Gv_{2,2}\Uv_{N^\prime}\mathbf{\Lambda}_{N^\prime}\Qv)
  -\frac{1}{T^\prime}h(\Gv_2\Uv_{N^\prime}\mathbf{\Lambda}_{N^\prime}\Qv)+d_{K,N_E,N_J,T^\prime,\text{partial-coh}}\nonumber\\
  &\leq&\frac{1}{T^\prime}\Big\{N_JN_E\log T^\prime-KN_J\log\pi eT^\prime\beta^2-N_E\sum_{i=1}^{N_J}\varphi(T^\prime-i+1)\log e\Big\}\nonumber\\
  &&+~d_{K,N_E,N_J,T^\prime,\text{partial-coh}},\nonumber\\
  c_{K,N_E,N_J,T^\prime,\text{partial-coh}}&\geq&\frac{1}{T^\prime}\Big\{(N_E-K)\sum_{i=1}^{N_J}\varphi(T^\prime-i+1)\log e-N_EN_J\log T^\prime-KN_J\log\pi e\beta^2\Big\}\nonumber\\
  &&+~d_{K,N_E,N_J,T^\prime,\text{partial-coh}},\nonumber
  \eea
  \bea
  \text{with}~~d_{K,N_E,N_J,T^\prime,\text{partial-coh}}\triangleq
  \Big(1-\frac{N_J}{T^\prime}\Big)\bigg(\sum_{i=1}^{N_J}\mathsf{E}\Big[\log\frac{\lambda_{G_{2,2}N^\prime,i}^2}{\lambda_{G_{2}N^\prime,i}^2}\Big]-K\log\pi e\bigg)
  ~~~~~~~~~~~~~~~~~~~\nonumber\\
  +\sum_{i=1}^K\varphi(N_E-i+1)\log e+K\log\pi e\alpha^2.\nonumber
\eea
\end{Theorem}

The result in Theorem \ref{Theorem:Theorem3} shows the exact DoF of the partially coherent leakage rate, from which we recognize that
the universal bound in Theorem \ref{Theorem:Theorem2} is tight in terms of the DoF
for the case of $N_E\geq\bar{M}$, but otherwise it may be loose especially at high $\text{SNR}_E$.
A zero-DoF condition $N_J\geq T^\prime$ arises from Theorem \ref{Theorem:Theorem3}.
Interestingly, for $T^\prime=T-K$, it is equivalently $\bar{M}\geq T$ which is the same condition for the non-coherent ED.
By combining the results in Theorem \ref{Theorem:Theorem2} and \ref{Theorem:Theorem3}, one may also find a tighter upperbound by taking the minimum of them,
but details are omitted for brevity.

\section{Discussion}\label{sec:Discussion}

In this section, we introduce useful design methods for the ANAM systems by leveraging the proposed analysis on the leakage rates.
To this end, we assume that the BS has an access at least to the statistical information of the ED channels such as the distribution, the ED's location, and the doppler frequency.
Note that the instantaneous CSIs of the ED channels are still unknown to the BS.

\subsection{How many antennas do we need at the BS for secrecy?}\label{sec:how many antennas}

In practice, a popular rule of thumb for determining the coherence time $T_c$ in modern digital communications is $T_c\simeq\frac{0.423}{f_m}$(sec)
where $f_m$(Hz) denotes the maximum doppler frequency.
In LTE-A, for example, one OFDM symbol duration is about $72.4\mu$s, which implies that
the coherence symbol length $T$ of the ED's channels in LTE-A is approximately computed by
\bea\label{eq:moving speed}
T=\frac{T_c}{72.4\mu\text{s}}=\frac{5842.5}{f_m},
\eea
where we have $f_m=\frac{vf_c}{c}$ for carrier frequency $f_c$(Hz), speed of a moving object $v$(m/s), and speed of light $c=3\times10^8$(m/s).

With the knowledge on the doppler frequency of the ED channels,
the BS may determine its number of antennas such that $M=\left\lceil\frac{5842.5}{f_m}\right\rceil$
to achieve the zero-DoF leakage rate.
Obviously as the ED's mobility $v$ or the carrier frequency $f_c$ increases,
the required number of BS antennas will decrease.
For example, for [$f_c=5$ GHz, $v=0.8$ m/s] and [$f_c=10$ GHz, $v=1.3$ m/s], the BS may need $M=351$ and $135$ number of antennas, respectively,
both of which are reasonable numbers from the massive MIMO point of view.\footnote{We note that in the practical wireless channels,
the coherence time should be finite even if the ED's mobility equals zero,
because the channel variation is also affected by the weather or moving obstacles around the ED.
It is also possible to intentionally change the ED's channel by installing shielding plates or moving obstacles around the BS antennas.
Such channel variations also can be effectively modeled by a doppler frequency of the ED channel.}

\subsection{Artificially fast faded AN}\label{sec:AFF}

When the coherence time of the ED channels is measured too long, we can deliberately randomize the artificial noise channel $\Gv_2$
by reformulating the transmit signal in (\ref{eq:transmit signal}) as
\bea\label{eq:AFF}
\Xv=\alpha\Hv^\dagger\Sv+\beta\bar{\Vv}_{H_2,N_J}\Av(t)\Nv,
\eea
where $\Av(t)$ denotes the AFF precoder that randomly changes in each time instant $t$ within the coherence time.
Then, the received signal at Eve can be expressed by
\bea\label{eq:AFF2}
\Yv_E=\Gv_1\Sv+\Gv_2(t)\Nv+\Zv,
\eea
where $\Gv_2(t)\triangleq\Gv_2\Av(t)$.
Observe that as the changing period of $\Av(t)$ becomes shorter, the effective coherence time of $\Gv_2(t)$ reduces, and therefore we obtain an enhanced security.

A major difference point of (\ref{eq:AFF}) from the existing AFF schemes in \cite{XLi:07} and \cite{YKozai:15} is that
we apply the AFF precoder to the artificial noise signal $\Nv$, not to the data signal $\Sv$.
The reason is that the AFF precoder applied to the data signal $\Sv$ may be cost ineffective
because it requires an additional effort to make the LU channels deterministic,
which reduces the randomness of the ED channels \cite{CSong:18TVT}.
In contrast, the AFF precoder in (\ref{eq:AFF}) does not require the channel constantization process,
because the AN will not appear in the LU channels.
In addition, as shown in Section \ref{sec:coherent leakage rate with partial CSI},
the short coherence time of $\Gv_2(t)$ may significantly degrade the ED's decoding ability even if $\Gv_1$ is perfectly known to the ED.
Therefore, the proposed AFF precoding in (\ref{eq:AFF}) is indeed effective.
Unfortunately, however, a rigorous analysis for the leakage-rate is unviable yet because independence between two fading channel matrices
$\Gv_2(t)$ and $\Gv_2(t^\prime)$ for $t\neq t^\prime$ is not guaranteed, and may be discussed in our future works.

\subsection{Achievable Secrecy Rate}

With a knowledge on the ED's channel statistics at the BS, it is also of interest to investigate the secrecy rate that specifies the secure data rate to the LUs with the aid of the wiretap codes at the BS.\footnote{In general, the wiretap code requires the BS to know the ED's CSI to compute the leakage rate $\mathcal{I}(\Sv;\Yv_E)$ since it is needed to determine the amount of the confusing codewords in each message bin \cite{MBloch:11} \cite{CLing:14}.
In our cases, the leakage rate can be evaluated based on the statistical information of the ED channels without needing to know their instantaneous values, and thus the secrecy-rate is also a valid metric.
The interested readers may refer to \cite{SGoel:08,HMWang:15,HSi:15} for other cases of the secrecy rate being investigated without the ED's CSI at the transmitter}
In general, the achievable secrecy rate is computed by the difference of the information rate to the LUs and the leakage rate to the ED for a given input distribution \cite{MBloch:11}. In what follows, we consider two scenarios, a single LU having $K$ antennas and $K$ LUs each having a single antenna.
Note that the achievable secrecy rates are obtained based on the leakage rates upperbounds in Theorems \ref{Theorem:Theorem1} and \ref{Theorem:Theorem3} rather than the lowerbounds.

\subsubsection{Single-user secrecy rate}
For the case of a single LU with $K$ antennas, the wiretap code can be applied across the $K$ different data streams, because we have only one message to encode.
Therefore, the achievable secrecy rate with non-coherent ED is simply given by
\bea\label{eq:SecrecyRate SU noncoherent}
R_{\text{sec,non-coh}}=\left[K\log\left(1+M\alpha^2\text{SNR}_L\right)-\mathcal{L}_{\text{non-coh}}(K,N_E,N_J,T)\right]^+.
\eea

As for the partially coherent ED, the data transmission takes place over $T^\prime$ symbol times during the coherent time $T$ due to the downlink training phase.
Therefore, the secrecy rate is computed as
\bea\label{eq:SecrecyRate SU coherent}
R_{\text{sec,partial-coh}}=\frac{T^\prime}{T}\left[K\log\left(1+M\alpha^2\text{SNR}_L\right)-\mathcal{L}_{\text{partial-coh}}(K,N_E,N_J,T^\prime)\right]^+.
\eea

\subsubsection{Multi-user secrecy rate} \label{eq:Multi-user secrecy rate}
For the case of multiple $K$ LUs with a single antenna, the wiretap code must be applied independently to each data stream, because otherwise the
encoded data may not be decodable at each LU.
A reasonable approach in this case is to consider the worst case scenario where the ED can intercept all other messages $\mathcal{M}_i,\forall i\neq k$
when we encode the message for the $k$-th user.
Thus, the achievable non-coherent and partially coherent secrecy-rates to the $K$ LUs are respectively computed as
\bea\label{eq:SecrecyRate MU noncoherent}
R_{\text{sec,non-coh}}&=&K\left[\log\left(1+M\alpha^2\text{SNR}_L\right)-\mathcal{L}_{\text{non-coh}}(1,N_E,N_J,T)\right]^+\nonumber\\
\label{eq:SecrecyRate MU coherent}
\text{and}~~R_{\text{sec,partial-coh}}&=&\frac{KT^\prime}{T}\left[\log\left(1+M\alpha^2\text{SNR}_L\right)-\mathcal{L}_{\text{partial-coh}}(1,N_E,N_J,T^\prime)\right]^+,
\eea

\section{Numerical Results}\label{sec:Simulation Results}

\begin{figure}
\begin{center}
\includegraphics[width=3.5in]{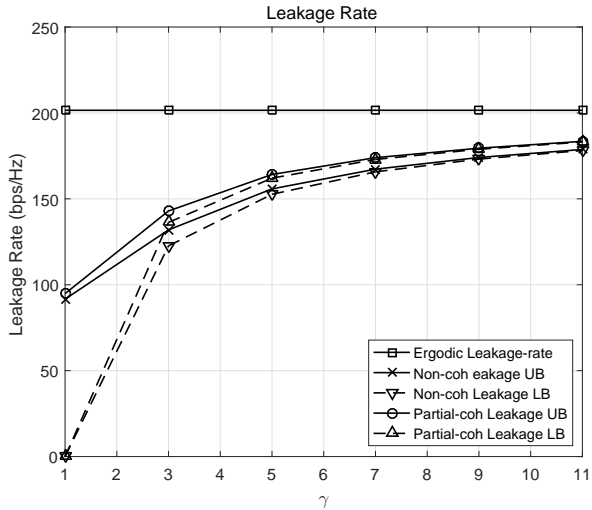}
\end{center}
\caption{Leakage rate performance according to various coherence time $T=\gamma M$ with $\text{SNR}_E=30$ dB. \label{Fig:figure1.eps}}
\end{figure}

In this section, we investigate the leakage and secrecy rate performance of the ANAM systems through numerical examples and
we provide useful observations.
Throughout the section, we focus on a system with $M=64$, $K=16$, and $\alpha=\beta=1$ unless stated otherwise,
but the result can be generally applied to other simulation environments.
As illustrated in Figure \ref{Fig:figure0.eps}, we assume that the distributions of the ED channels $\Gv_1$ and $\Gv_2$ follow Properties \ref{Property:Property1} and \ref{Property:Property2}.
In addition, the effective coherence time for the partially coherent scenario equals $T^\prime=T-K$.

Figure \ref{Fig:figure1.eps} compares the leakage rate performance for various CSI situations at the ED in the ANAM systems with
$N_E=64$, $N_J=48$, $\text{SNR}_E=30$ dB, and $T=\gamma M$ for $\gamma\geq1$.
Here, ``{\it Ergodic Leakage}'' indicates the ergodic leakage rate in Lemma \ref{Lemma:Lemma1} for the coherent ED, which amounts to the worst case in terms of security,
and ``{\it UB}'' and ``{\it LB}'' denote the proposed upper and lower bounds of the leakage rates.
It is seen that as the coherence time $T$ increases, both the partially-coherent and non-coherent leakage rates approach their full CSI counter part.
This is because a large coherence time may give enough time for the ED to perform the blind detection \cite{MFrikel:00} \cite{YChi:10}.
The figure also confirms our statement in equation (\ref{eq:entropy gap}) that
as the coherence time increases, the gap between the leakage rate bounds reduces.
In most cases, the bounds are tight when $T>5M$, at which we can estimate the exact amount of leakage to the ED.
The partially coherent leakage is slightly higher than the non-coherent one due to $\Gv_1$ known at ED.

\begin{figure}
\begin{center}
\includegraphics[width=3.5in]{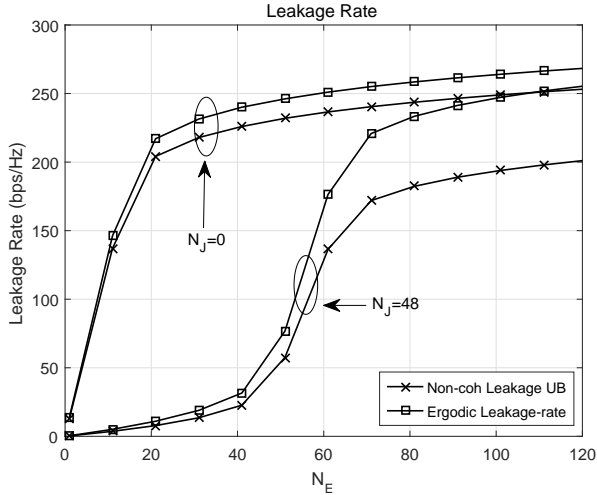}
\end{center}
\caption{Leakage rate performance according to various $N_E$ and $N_J$ with $T=5M$ and $\text{SNR}_E=30$ dB. \label{Fig:figure2.eps}}
\end{figure}

Figure \ref{Fig:figure2.eps} shows the leakage rate performance according to various $N_E$ and $N_J$ with $T=5M$ and $\text{SNR}_E=30$ dB.
Note that for the case of $N_J=0$, we set $\alpha=\sqrt{M/K}$.
As observed in \cite{DKapetanovic:15}, when $M\gg N_E$, the ED's decoding ability keeps small regardless of the AN and CSI at the ED.
However, as $N_E$ gets larger to be compared with $M$, the AN becomes essential to disrupt the ED.
One interesting observation is that the coherent ED manages the AN better than the non-coherent ED especially when $N_E>M$,
which verifies the robustness of the ANAM system to the large antenna array attacks from the ED.

\begin{figure}
\begin{center}
\includegraphics[width=3.5in]{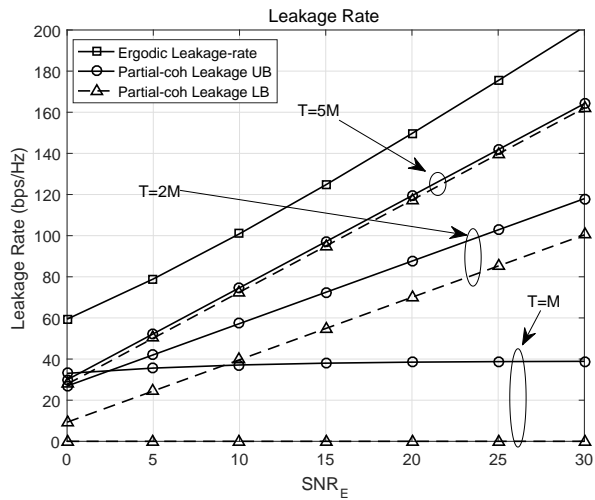}
\end{center}
\caption{Leakage rate performance according to various $\text{SNR}_E$ and $T$. \label{Fig:figure3.eps}}
\end{figure}

Figure \ref{Fig:figure3.eps} exhibits the partially coherent leakage-rate bounds according to various $\text{SNR}_E$ and $T$ with $N_E=64$ and $N_J=48$.
First, we verify that the AN channel uncertainty at the ED degrades its DoF by the factor of $(1-N_J/T^\prime)^+$ compared to the ergodic leakage rate.
Also, it is confirmed again that the proposed bounds are tight over the $\text{SNR}$s for $T>5M$.
In contrast, if $T$ keeps decreasing towards $M$, the bounds may be loose.
Nevertheless, the information leakage effect will be reduced at high SNR,
because in this case the leakage-rate DoF gradually disappears.

\begin{figure}
\begin{center}
\includegraphics[width=3.5in]{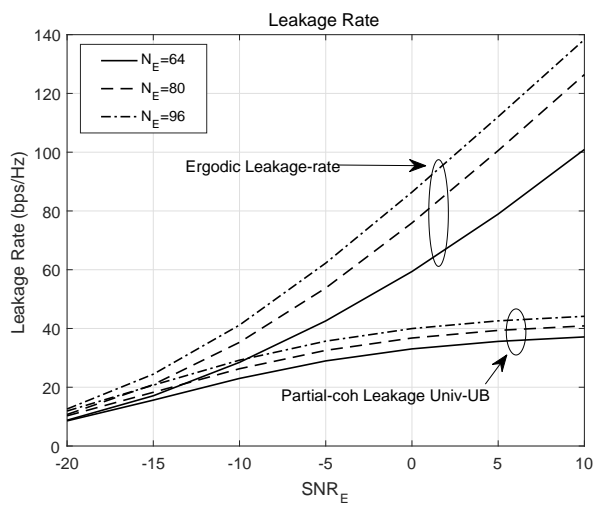}
\end{center}
\caption{Leakage rate performance in the low $\text{SNR}_E$ regime with $T=M$ and various $N_E$. \label{Fig:figure4.eps}}
\end{figure}

Figure \ref{Fig:figure4.eps} shows the low SNR behavior of the universal leakage-rate upperbound with $N_J=48$, $T=M$, and $N_E=64$.
We confirm that when $\text{SNR}_E$ is sufficiently small, the universal bound exhibits the same behavior as the ergodic leakage rate.
This is because the AN is relatively ignorable at low $\text{SNR}_E$.
On the contrary, as $\text{SNR}_E$ increases, the universal bound will be saturated by means of the AN and ED's channel uncertainty on $\Gv_2$.
Note that the performance of the universal bound presented in this figure is also helpful to predict
the low SNR behavior of the non-coherent leakage rates as well as the partially coherent one.

\begin{figure}
\begin{center}
\includegraphics[width=3.5in]{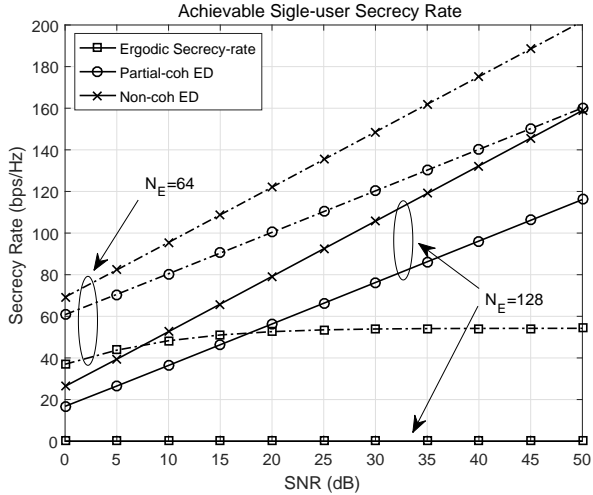}
\end{center}
\caption{Secrecy rate performance towards a single LU with $K$ antennas according to $\text{SNR}_E=\text{SNR}_L=\text{SNR}$ and $N_E$ with $T=2M$.  \label{Fig:figure5.eps}}
\end{figure}

Figure \ref{Fig:figure5.eps} illustrates the achievable secrecy rate to a single LU with $K$ antennas in the ANAM system
with $N_J=48$, $T=2M$, and $\text{SNR}_E=\text{SNR}_L=\text{SNR}$.
For all $N_E$, the LUs with the non-coherent ED can enjoy higher secrecy rate over those with the partially coherent ED,
because the former does not pay for the downlink training.
On the other hand, the LUs with the partially coherent ED achieves an improved secrecy-rate compared to the case of the coherent ED
due to the unknown CSI on $\Gv_2$ and the limited coherence time $T$.
This confirms our statement in Section \ref{sec:AFF} that
the AFF scheme which randomizes $\Gv_2$ will further mitigates the leakage-rate by reducing the effective coherence time $T$.
It is also interesting to observe that both the partially coherent and non-coherent secrecy-rates continuously grow with $\text{SNR}$ regardless of $N_E$,
whereas the ergodic secrecy-rate even vanishes as $N_E$ increases from $64$ to $128$.

\begin{figure}
\begin{center}
\includegraphics[width=3.5in]{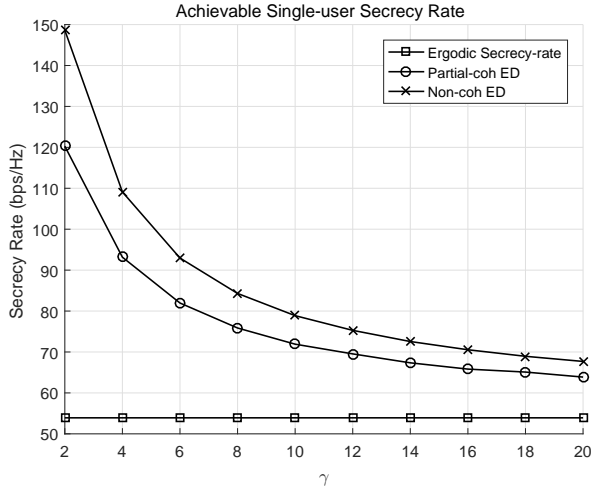}
\end{center}
\caption{Secrecy rate performance towards a single LU with $K$ antennas according to the coherence time $T=\gamma M$
and $\text{SNR}_E=\text{SNR}_L=30$ dB.  \label{Fig:figure6.eps}}
\end{figure}
\begin{figure}
\begin{center}
\includegraphics[width=3.5in]{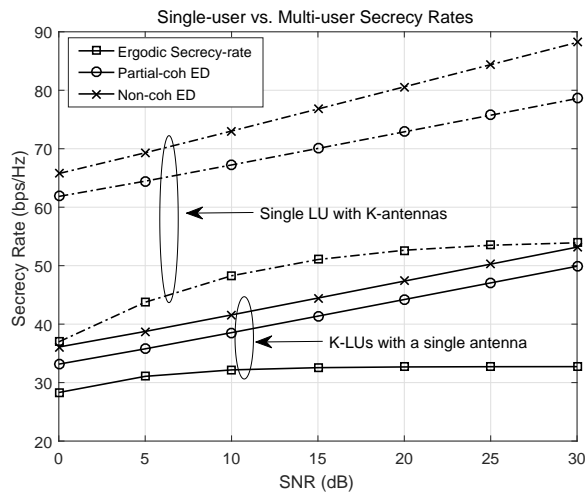}
\end{center}
\caption{Single user secrecy-rate vs. multi-user secrecy-rate according to $\text{SNR}_E=\text{SNR}_L=\text{SNR}$ with $T=7M$. \label{Fig:figure7.eps}}
\end{figure}

In Figure \ref{Fig:figure6.eps}, we investigate the tendency of the secrecy-rate change according to the coherence time $T=\gamma M$ with $N_J=48$, $\text{SNR}_E=\text{SNR}_L=30$ dB.
We observe that the smaller the ED's channel coherence time is, the higher the achievable secrecy-rate becomes.
As illustrated in Section \ref{sec:how many antennas}, the coherent time $T=2M$ amounts to the ED's mobility $1.3$ km/h in the carrier frequency of $f_c=10$ GHz in LTE-A, which implies that
a small movement of the ED or a slight environmental change around the ED may produce a high security gain in the ANAM systems.

Figure \ref{Fig:figure7.eps} compares the performance of single-user and multi-user secrecy rates with $N_J=48$, $T=7M$, and $\text{SNR}_E=\text{SNR}_L$.
As explained in Section \ref{eq:Multi-user secrecy rate}, the single-user secrecy-rate outperforms the multi-user secrecy-rate, because each encoding function for multiple LUs
considers a genie aided ED that is more capable than the one for the single-user case.

\section{Conclusion}\label{sec:Conclusion}

In this paper, we have investigated the leakage-rate performance of the ANAM systems with a non-coherent passive ED in block fading,
where CSI of the ED channels are unknown to both the BS and the ED.
First, we analyzed tight upper and lower bounds of the non-coherent leakage rate with the high SNR approximation.
Then, we derived the leakage rate to the coherent ED with partial CSI considering a situation in which the BS transmits the precoded downlink training.
We also computed single- and multi-user secrecy rates that are attainable via appropriate wiretap codes at the BS.
From our analysis, it was found that the conventional assumption on the full CSI ED may be too pessimistic to provide meaningful information on the security,
because the channel uncertainty at the ED may significantly degrade the ED's message interception ability.
Finally, the numerical results demonstrated that our analysis is accurate and the derived bounds are tight
as the coherence time is relatively larger than the number of BS antennas.

\appendix

\subsection{Proof of Proposition \ref{Proposition:Proposition1}}\label{Apx:Appendix A}

Note that in this proof, we basically assume that $M\leq T$. Let us define SVD of $\Yv_E=\Uv_E\mathbf{\Lambda}_E\Vv_E^H$ where
$\Uv_E\in\mathbb{C}^{N_E\times\Omega}$ and $\Vv_E\in\mathbb{C}^{T\times\Omega}$ are unitary matrices and
$\mathbf{\Lambda}_E\in\mathbb{C}^{\Omega\times\Omega}$ denotes a diagonal matrix having $\Omega\triangleq\min(N_E,T)$
non-zero singular values of $\Yv_E$ ordered in descending order, i.e., $\{\sigma_{1},\ldots,\sigma_{\Omega}\}$ on its main diagonal.
Also, observe that $\Yv_E$ is i.d., since it's distribution is invariant over both left and right unitary transformations,
which means that the singular vectors of $\Yv_E$ are i.d and independent of the singular values.
Thus, by the SVD coordinate change \cite{TAnderson:58}, we write
\bea
h(\Yv_E)&=&h(\Uv_E)+h(\Vv_E)+h(\sigma_{1},\ldots,\sigma_{\Omega})+\mathsf{E}[\log|J_{\Omega^\prime,\Omega}(\sigma_{1},\ldots,\sigma_{\Omega})|]\nonumber\\
&=&\log|S(N_E,\Omega)|+\log|S(T,\Omega)|+h(\sigma_{1},\ldots,\sigma_{\Omega})+\mathsf{E}[\log|J_{\Omega^\prime,\Omega}(\sigma_{1},\ldots,\sigma_{\Omega})|],\nonumber
\eea
where $\Omega^\prime\triangleq\max(T,N_E)$ and $J_{\Omega^\prime,\Omega}(\cdot)$ denotes
the Jacobian that is induced by the change of SVD coordinate,
which is defined by
\bea
J_{\Omega^\prime,\Omega}(\sigma_{1},\ldots,\sigma_{\Omega})\triangleq\prod_{i<j}(\sigma_{i}^2-\sigma_{j}^2)^2\prod_{i=1}^\Omega\sigma_{i}^{2(\Omega^\prime-\Omega)+1}.\nonumber
\eea

Now, the remaining problem is to find the differential entropy of the singular values of $\Yv_E$.
To this end, we first introduce the following lemma.
\begin{Lemma}\label{Lemma:Lemma3}
For the singular values of $\Yv_E$ with $\bar{\Gv}\bar{\Xv}$ having rank $\Xi\triangleq\min(N_E,\bar{M})$, the following property holds as $\sigma_z^2\rightarrow0$,
\bea
\{\sigma_{1},\ldots,\sigma_{\Xi}\}&\overset{d}{\rightarrow}&\{\mu_{1},\ldots,\mu_{\Xi}\}\nonumber\\
\text{and}~~\{\sigma_{\Xi+1},\ldots,\sigma_{\Omega}\}&\overset{d}{\rightarrow}&\{\mu_{\Xi+1},\ldots,\mu_{\Omega}\}\nonumber
\eea
where $\{\mu_{1},\ldots,\mu_{\Xi}\}$ and $\{\mu_{\Xi+1},\ldots,\mu_{\Omega}\}$ are the singular values of $\bar{\Gv}\bar{\Xv}$
and an independent $(N_E-\Xi)\times(T-\Xi)$ matrix $\bar{\Zv}$ having i.i.d. $\mathcal{CN}(0,\sigma_z^2)$ entries,
respectively.
\end{Lemma}
\begin{IEEEproof}
The proof is similar to \cite[Lemma 16]{LZheng:02}. However, the result in \cite[Lemma 16]{LZheng:02} is restricted to the cases of $N_E>\bar{M}$ and a specific form of input signal $\bar{\Xv}$.
Here, we provide a refined proof that holds for general $N_E$ and any matrix $\bar{\Gv}\bar{\Xv}$ that is independent of $\Zv$.

By the circular symmetry of the noise matrix $\Zv$, the singular values of $\Yv_E$ has the same distribution with those of
\bea
\Yv_0=\bigg[\begin{array}{cc}
              \mathbf{D} & \mathbf{0} \\
              \mathbf{0} & \mathbf{0}
            \end{array}
\bigg]+
\bigg[\begin{array}{cc}
        \Zv_{11} & \Zv_{11} \\
        \Zv_{21} & \Zv_{22}
      \end{array}
\bigg]\nonumber
\eea
where $\Dv\in\mathbb{R}^{\Xi\times\Xi}$ denotes the non-zero singular value matrix of $\bar{\Gv}\bar{\Xv}$, and
$\Zv_1\in\mathbb{C}^{\Xi\times\Xi}$, $\Zv_2\in\mathbb{C}^{\Xi\times(T-\Xi)}$, and $\Zv_3\in\mathbb{C}^{(N_E-\Xi)\times(T-\Xi)}$
represents random matrices having i.i.d. $\mathcal{CN}(0,\sigma_z^2)$ entries.
Now, consider an equation $f(\lambda)=\det(\lambda\Iv_{N_E}-\Yv_0\Yv_0^\mathsf{H})$ with the roots $(\sigma_1^2,\ldots,\sigma_{N_E}^2)$.
Then, it can be simplified at high $\text{SNR}_E$ as
\bea\label{eq:determinant}
f(\lambda)&\overset{\sigma\rightarrow0}{=}&\det\Bigg(\bigg[\begin{array}{cc}
              \lambda\Iv_{\Xi}-\Dv^2 & -\Dv\Zv_{21}^\mathsf{H} \\
              -\Zv_{21}^\mathsf{H}\Dv & \lambda\Iv_{N_E-\Xi}-\Zv_{21}\Zv_{21}^\mathsf{H}-\Zv_{22}\Zv_{22}^\mathsf{H}
            \end{array}
\bigg]\Bigg)\nonumber\\
&=&\det(\lambda\Iv_{\Xi}-\Dv^2)\det(\lambda_{N_E-\Xi}-\Zv_{22}\Zv_{22}^\mathsf{H}-\Zv_{21}\Zv_{21}^\mathsf{H}+\Zv_{21}\Dv(\Dv^2-\lambda_{\Xi})^{-1}\Dv\Zv_{21}^\mathsf{H},
\eea
where the second equality follows from the Schur's identity for determinant of a block matrix.
To find the roots of $f(\lambda)=0$, we observe that the first $\Xi$ roots are the entries in $\Dv^2$.
The remaining $N_E-\Xi$ eigenvalues are thus order of $\sigma_z^2$, which means that they are much smaller than the entries of $\Dv$.
We can approximate ($\Dv^2-\lambda\Iv_{\Xi}$) as $\Dv^2$, i.e., the second determinant of (\ref{eq:determinant}) becomes
$\det(\lambda\Iv_{\Xi-N_E}-\Zv_{22}\Zv_{22}^\mathsf{H})$. Therefore, the remaining $N_E-\Xi$ eigenvalues of $\Yv_0\Yv_0$ are approximately the eigenvalues of $\Zv_{22}\Zv_{22}^\mathsf{H}$,
and the proof is completed.
\end{IEEEproof}

Lemma \ref{Lemma:Lemma3} states that the two sets of singular values $\{\sigma_{1},\ldots,\sigma_{\Xi}\}$ and $\{\sigma_{\Xi+1},\ldots,\sigma_{\Omega}\}$
of $\Yv_E$ are asymptotically independent of each other.
Thus, we have
\bea\label{eq:proof of Theorem 1-1}
h(\Yv_E)&=&\log|S(N_E,\Omega)|+\log|S(T,\Omega)|+h(\sigma_{1},\ldots,\sigma_{\Xi})+h(\sigma_{\Xi+1},\ldots,\sigma_{\Omega})\nonumber\\
&&+\mathsf{E}[\log|J_{\Omega^\prime,\Omega}(\sigma_{1},\ldots,\sigma_{\Omega})|].
\eea
Note that the singular values of $\bar{\Gv}\bar{\Xv}$ and $\hat{\Gv}\mathbf{\Lambda}_{\bar{X}}$ are the same.
Also, a matrix $\hat{\Gv}\mathbf{\Lambda}_{\bar{X}}\Qv$ is i.d., since $\Qv$ is independent of $\bar{\Gv}\bar{\Xv}$.
Thus, we can consider the following differential entropy via the SVD coordinate change as
\bea\label{eq:proof of Theorem 1-2}
h(\hat{\Gv}\mathbf{\Lambda}_{\bar{X}}\Qv)&=&\log|S(N_E,\Xi)|+\log|S(\bar{M},\Xi)|+h(\sigma_{1},\ldots,\sigma_{\Xi})\nonumber\\
&&+\mathsf{E}[\log|J_{\Xi^\prime,\Xi}(\sigma_{1},\ldots,\sigma_{\Xi})|],
\eea
where $\Xi^\prime\triangleq\max(N_E,\bar{M})$.
Also, we write
\bea\label{eq:proof of Theorem 1-3}
h(\bar{\Zv})&=&(N_E-\Xi)^+(T-\Xi)\log\pi e \sigma_z^2\nonumber\\
&=&\log|S(\Omega-\Xi,\Omega-\Xi)|+\log|S(\Omega^\prime-\Xi,\Omega-\Xi)|+h(\sigma_{\Xi+1},\ldots,\sigma_{\Omega})\nonumber\\
&&+\mathsf{E}[\log|J_{\Omega^\prime-\Xi,\Omega-\Xi}(\sigma_{\Xi+1},\ldots,\sigma_{\Omega})|].
\eea
where the second equality follows from the SVD coordinate change of $\bar{\Zv}$.
Then, combining the three equations from (\ref{eq:proof of Theorem 1-1}) to (\ref{eq:proof of Theorem 1-3}), we get
\bea
h(\Yv_E)&=&h(\hat{\Gv}\mathbf{\Lambda}_{\bar{X}}\Qv)+(N_E-\Xi)^+(T-\Xi)\log\pi e \sigma_z^2\nonumber\\
&&+\mathsf{E}[\log|J_{\Omega^\prime,\Omega}(\sigma_{1},\ldots,\sigma_{\Omega})|]-\mathsf{E}[\log|J_{\Xi^\prime,\Xi}(\sigma_{1},\ldots,\sigma_{\Xi})|]\nonumber\\
&&-\mathsf{E}[\log|J_{\Omega^\prime-\Xi,\Omega-\Xi}(\sigma_{\Xi+1},\ldots,\sigma_{\Omega})|]\nonumber\\
&&+\log|S(N_E,\Omega)|+\log|S(T,\Omega)|-\log|S(N_E,\Xi)|-\log|S(\bar{M},\Xi)|\nonumber\\
&&-\log|S(\Omega-\Xi,\Omega-\Xi)|-\log|S(\Omega^\prime-\Xi,\Omega-\Xi)|.
\eea

First, for $N_E\leq\bar{M}$, it follows that
\bea
h(\Yv_E)=h(\hat{\Gv}\mathbf{\Lambda}_{\bar{X}}\Qv)+\mathsf{E}[\log|J_{T,N_E}(\sigma_{1},\ldots,\sigma_{N_E})|]
-\mathsf{E}[\log|J_{\bar{M},N_E}(\sigma_{1},\ldots,\sigma_{N_E})|]~~~\nonumber\\
+\log|S(T,N_E)|-\log|S(\bar{M},N_E)|~~~~~~~~~~~~~~~~~~~~~~~~~~~~~~~~~~~~~~~~~~~~~~~\nonumber\\
=h(\hat{\Gv}\mathbf{\Lambda}_{\bar{X}}\Qv)+\log\frac{|G(T,N_E)|}{|G(\bar{M},N_E)|}
+\sum_{i=1}^{N_E}\mathsf{E}\left[\log\sigma_i^{2(T-N_E)+1}-\log\sigma_i^{2(\bar{M}-N_E)+1}\right]\nonumber
\eea
\bea\label{eq:proof of Theorem 1-4}
=h(\hat{\Gv}\mathbf{\Lambda}_{\bar{X}}\Qv)+\log\frac{|G(T,N_E)|}{|G(\bar{M},N_E)|}+(T-\bar{M})\mathsf{E}\left[\sum_{i=1}^{N_E}\log\sigma_i^{2}\right].
~~~~~~~~~~~~~~~~~~~
\eea
Similarly, for $N_E>\bar{M}$, we can show that
\bea\label{eq:proof of Theorem 1-5}
h(\Yv_E)&=&h(\hat{\Gv}\mathbf{\Lambda}_{\bar{X}}\Qv)+(N_E-\bar{M})^+(T-\bar{M})\log\pi e \sigma_z^2\nonumber\\
&&+\log|G(T,\bar{M})|+(T-\bar{M})\mathsf{E}\left[\sum_{i=1}^{\bar{M}}\log\sigma_i^{2}\right].
\eea
Finally, combining (\ref{eq:proof of Theorem 1-4}) and (\ref{eq:proof of Theorem 1-5}), we obtain Theorem \ref{Proposition:Proposition1}.

\subsection{Proof of Lemma \ref{Lemma:Lemma2}}\label{Apx:Appendix B}
By Properties \ref{Property:Property1} and \ref{Property:Property2}, we have
$\mathsf{E}[\text{Tr}(\bar{\Gv}\Uv_{\bar{X}}\mathbf{\Lambda}_{\bar{X}}^2\Uv_{\bar{X}}^\mathsf{H}\bar{\Gv}^H]
=\mathsf{E}[\text{Tr}(\bar{\Gv}\bar{\Xv}\bar{\Xv}^\mathsf{H}\bar{\Gv}^\mathsf{H})]
=TN_E(\alpha^2 K+\beta^2 N_J)
=TN_EM$
where the last equality stems from the power constraint in (\ref{eq:power constraint approx}). Thus,
the differential entropy $h(\hat{\Gv}\mathbf{\Lambda}_{\bar{X}}\Qv)$ is maximized by
the $N_E\times \bar{M}$ matrix with i.i.d. $\mathcal{CN}(0,T)$ entries as
$h(\hat{\Gv}\mathbf{\Lambda}_{\bar{X}}\Qv)\leq \bar{M}N_E\log\pi e T$.

Now, the lower bound can be computed as
\bea
h(\hat{\Gv}\mathbf{\Lambda}_{\bar{X}}\Qv)&\geq& h(\bar{\Gv}\Uv_{\bar{X}}\mathbf{\Lambda}_{\bar{X}}\Qv|\bar{\Xv},\Qv)\nonumber\\
&=&h(\bar{\Gv})+N_E\mathsf{E}\left[\log\det\left(\bar{\Xv}\bar{\Xv}^\mathsf{H}\right)\right]\nonumber\\
&=&N_E(K\log\pi e\alpha^2+N_J\log\pi e\beta^2)+N_E\sum_{i=1}^{\bar{M}}\varphi(T-i+1)\log e\nonumber
\eea
where the last equality follows from Definition \ref{Def:Definition3}, and the proof is completed.

\subsection{Proof of Theorem \ref{Theorem:Theorem2}}\label{Apx:Appendix C}

For the proof, we first introduce the following lemma.
\begin{Lemma}\label{Lemma:Lemma4}
For the given $\Yv_E^\prime$ in (\ref{eq:leakage partialCSI}), we obtain the following inequality
\bea
\mathcal{I}(\Yv_E^\prime;\Nv^\prime|\Gv_1)<\mathcal{I}(\Yv_E^\prime;\Nv^\prime|\Gv_1,\Sv^\prime)
\eea
which is tight as $\sigma_z^2\rightarrow\infty$.
\end{Lemma}
\begin{IEEEproof}
First, observe that $\mathcal{I}(\Yv_E^\prime;\Nv^\prime|\Gv_1)$ amounts to a fictitious case of the BS sending the signal $\Nv$
to the ED with the effective additive noise $\Gv_1\Sv^\prime+\Zv^\prime$ where $\Gv_1$ is known to the ED.
Therefore, the leakage rate will further increase if the ED knows both $\Gv_1$ and $\Sv^\prime$ because in this case the noise reduces to $\Zv$, which is the case of
$\mathcal{I}(\Yv_E^\prime;\Nv^\prime|\Gv_1,\Sv^\prime)$.
When $\sigma_z\rightarrow\infty$, $\Gv_1\Sv^\prime$ is ignorable relative to $\Zv$, which implies that
$\mathcal{I}(\Yv_E^\prime;\Nv^\prime|\Gv_1)\simeq\mathcal{I}(\Yv_E^\prime;\Nv^\prime|\Gv_1,\Sv^\prime)$.
Therefore, the bound in (\ref{eq:leakage upper partialCSI}) is tight in the low $\text{SNR}_E$ regime.
\end{IEEEproof}

Now, based on Lemma \ref{Lemma:Lemma4}, one can verify the following inequality as
\bea\label{eq:leakage upper partialCSI}
\mathcal{I}(\Yv_E^\prime;\Sv^\prime|\Gv_1)&=&\mathcal{I}(\Yv_E^\prime;\Sv^\prime,\Gv_2,\Nv^\prime|\Gv_1)-\mathcal{I}(\Yv_E^\prime;\Gv_2,\Nv^\prime|\Gv_1,\Sv^\prime)\nonumber\\
&=&\mathcal{I}(\Yv_E^\prime;\Sv^\prime,\Gv_2,|\Gv_1,\Nv^\prime)-\mathcal{I}(\Yv_E^\prime;\Gv_2|\Nv^\prime,\Gv_1,\Sv^\prime)\nonumber\\
&&+\mathcal{I}(\Yv_E^\prime;\Nv^\prime|\Gv_1)-\mathcal{I}(\Yv_E^\prime;\Nv^\prime|\Gv_1,\Sv^\prime)\nonumber\\
&\leq&\mathcal{I}(\Yv_E^\prime;\Sv^\prime,\Gv_2,|\Gv_1,\Nv^\prime)-\mathcal{I}(\Yv_E^\prime;\Gv_2|\Nv^\prime,\Gv_1,\Sv^\prime),
\eea
where the first two equalities follow from the chain rules and the last inequality stems from Lemma \ref{Lemma:Lemma4}, and thus is tight at low $\text{SNR}_E$.

The right-hand side of (\ref{eq:leakage upper partialCSI}) can be evaluated as in the following.
First, we consider that
\bea\label{eq:lemma2-1}
\mathcal{I}(\Yv_E^\prime;\Gv_2|\Nv^\prime,\Gv_1,\Sv^\prime)=\mathcal{I}(\Nv^{\prime\mathsf{T}}\Gv_2^{\mathsf{T}}+\Zv^{\prime\mathsf{T}};\Gv_2^{\mathsf{T}}|\Nv^\prime)~~~~~~~~~~~~~~~~~~~~~~~~~~~~~~~~~~~~~~~~~~~~~~\\
\label{eq:lemma2-2}
=N_E\mathsf{E}\left[\log\det\left(\beta^2\Nv^{\prime\mathsf{T}}\Nv^{\prime*}+\sigma_z^2\Iv_{T^\prime}\right)\right]-N_ET^\prime\log\sigma_z^2~~~~~~~~~~~~~~~~~\\
\label{eq:lemma2-3}
=N_E\min(N_J,T^\prime)\log\text{SNR}_E+N_E\mathsf{E}\bigg[\sum_{i=1}^{\min(N_J,T^\prime)}\log(\beta^2\lambda_{N^\prime,i}^2+\sigma_z^2)\bigg]
\eea
where the second equality is due to the fact that (\ref{eq:lemma2-1})
represents a virtual MIMO channel where a transmitter with $N_J$ antennas sends the i.i.d. $\mathcal{CN}(0,\beta^2)$ signals in $\Gv_2^{\mathsf{T}}$ during $N_E$ symbol times
through a random propagation matrix $\Nv^{\prime\mathsf{T}}$ that is known to the ED.

Next, the following properties are useful for further derivations.
\begin{Property}\label{Property:Property3}
  As for the Kronecker product, the following properties hold $(\Av\otimes\Bv)(\Cv\otimes\Dv)=\Av\Cv\otimes\Bv\Dv$ and $(\Av\otimes\Bv)^\mathsf{H}=\Av^\mathsf{H}\otimes\Bv^\mathsf{H}$.
\end{Property}
\begin{Property}\label{Property:Property4}
  For two square matrices $\Av\in\mathbb{C}^{p\times p}$ and $\Bv\in\mathbb{C}^{q\times q}$, the Kronecker sum is defined as
  $\Av\oplus\Bv\triangleq\Av\otimes\Iv_q+\Iv_p\otimes\Bv$. Then, the $(iq+j)$-th eigenvalue of $\Av\oplus\Bv$ equals $a_i^2+b_j^2$
  where $a_i^2$ and $b_j^2$ denote the $i$-th and $j$-th eigenvalues of $\Av$ and $\Bv$, respectively.
\end{Property}

Now, to evaluate $\mathcal{I}(\Yv_E^\prime;\Sv^\prime,\Gv_2,|\Gv_1,\Nv^\prime)$ in (\ref{eq:leakage upper partialCSI}),
we write $\Yv_E^\prime$ in the vectorization form as
\bea
\text{vec}(\Yv_E^\prime)&=&(\Iv_{T}\otimes\Gv_1)\text{vec}(\Sv^\prime)+(\alpha\Nv^{\prime\mathsf{T}}\otimes\Iv_{N_E})\text{vec}(\Gv_2/\alpha)+\text{vec}(\Zv^\prime)\nonumber\\
&=&[\Iv_T\otimes\Gv_1~~\beta\Nv^{\prime\mathsf{T}}\otimes\Iv_{N_E}]\nonumber
\left[\begin{array}{c}
\text{vec}(\Sv^\prime) \\
\text{vec}(\Gv_2/\beta)
\end{array}\right]+\text{vec}(\Zv^\prime)
\eea
where the components of $\Sv^\prime$ and $\Gv_2/\beta$ are i.i.d. $\mathcal{CN}(0,1)$.
Then, utilizing the two properties above, we can verify the following equivalences
\bea
\mathcal{I}(\Yv_E^\prime;\Sv^\prime,\Gv_2|\Gv_1,\Nv^\prime)=\mathcal{I}(\text{vec}(\Yv_E^\prime);\text{vec}(\Sv^\prime),\text{vec}(\Gv_2/\beta)|\Gv_1,\Nv^\prime)~~~~~~~~~~~~~~~~~~~~~~~~~~~~\nonumber\\
=\mathsf{E}\left[\log\det\left((\Iv_{T^\prime}\otimes\Gv_1\Gv_1^\mathsf{H})+(\beta^2\Nv^{\prime\mathsf{T}}\Nv^{\prime*}\otimes\Iv_{N_E})
+\sigma_z^2\Iv_{N_ET^\prime}\right)\right]-N_ET^\prime\log\sigma_z^2\nonumber\\
=\mathsf{E}\left[\log\det\left((\beta^2\Nv^{\prime\mathsf{T}}\Nv^{\prime*}\oplus\Gv_1\Gv_1^\mathsf{H})+\sigma_z^2\Iv_{N_ET^\prime}\right)\right]-N_ET^\prime\log\sigma_z^2.~~~~~~~~~~~~~~~~~
\eea
The rank of a kronecker sum $\beta^2\Nv^\mathsf{T}\Nv^*\oplus\Gv_1\Gv_1^\mathsf{H}$
equals $N_ET^\prime-R$ from Property \ref{Property:Property4} where $R\triangleq (T^\prime-N_J)^+(N_E-K)^+$ denotes the number of zero eigenvalues. Thus, we have
\bea\label{eq:lemma2-4}
\mathcal{I}(\Yv_E^\prime;\Sv^\prime,\Gv_2|\Gv_1,\Nv^\prime)~~~~~~~~~~~~~~~~~~~~~~~~~~~~~~~~~~~~~~~~~~~~~~~~~~~~~~~~~~~~~~~~~~~~~~~~~~~~~~~~~~\nonumber\\
=(N_ET^\prime-R)\log\text{SNR}_E+\mathsf{E}\bigg[\sum_{i=1}^{\min(T^\prime,N_J)}\sum_{j=1}^{\min(N_E,K)}\log(\beta^2\lambda_{N^\prime,i}^2+\lambda_{G_1,j}^2+\sigma_z^2)\bigg]~~~~~~~~~~~~~~~~\nonumber\\
+(N_E-K)^+\mathsf{E}\bigg[\sum_{i=1}^{\min(T^\prime,N_J)}\!\!\log(\beta^2\lambda_{N^\prime,i}^2+\sigma_z^2)\bigg]
\!\!+(T^\prime-N_J)^+\mathsf{E}\bigg[\sum_{j=1}^{\min(N_E,K)}\log(\lambda_{G_1,j}^2+\sigma_z^2)\bigg].
\eea
Finally, combining the results in (\ref{eq:lemma2-3}) and (\ref{eq:lemma2-4}), we obtain the theorem.

\bibliographystyle{ieeetr}
\bibliography{AZREF}

\begin{thebibliography}{10}

\bibitem{CLing:13}
C.~Ling, ``{Physec concepts for wireless public networks - Introduction, state
  of the art and perspectives},'' in {\em Proc. SDR Winncomm 2013}, Jan. 2013.

\bibitem{SGoel:08}
S.~Goel and R.~Negi, ``{Guaranteeing secrecy using artificial noise},'' {\em
  IEEE Trans. Wireless Commun.}, vol.~7, pp.~2180--2189, Jun. 2008.

\bibitem{AKhisti:10}
A.~Khisti and G.~W. Wornell, ``{Secure transmission with multiple antennas -
  PartI: The MISOME wiretap channel},'' {\em IEEE Trans. Inf. Theory}, vol.~56,
  pp.~3088 -- 3104, Jul. 2010.

\bibitem{PHLin:13}
P.~H. Lin, S.~H. Lai, S.~C. Lin, and H.~J. Su, ``{On secrecy rate of the
  generalized artificial-noise assisted secure beamforming for wiretap
  channels},'' {\em IEEE J. Sel. Areas Commun.}, vol.~31, pp.~1728--1740, Sep.
  2013.

\bibitem{HMWang:15a}
H.~M. Wang, C.~Wang, and D.~W.~K. Ng, ``{Artificial noise assisted secure
  transmission under training and feedback},'' {\em IEEE Trans. Sig. Process.},
  vol.~63, pp.~6285--6298, Dec. 2015.

\bibitem{TYLiu:17}
T.~Y. Liu, S.~C. Lin, and Y.~W.~P. Hong, ``{On the role of artificial noise in
  training and data transmission for secret communications},'' {\em IEEE Trans.
  Inf. Forensic and Secur.}, vol.~12, pp.~516--531, Mar. 2017.

\bibitem{DKapetanovic:15}
D.~Kapetanovic, G.~Zheng, and F.~Rusek, ``{Physical layer security for massive
  MIMO: An overview on passive eavesdropping and active attack},'' {\em IEEE
  Commun. Mag.}, vol.~53, pp.~21--27, Jun. 2015.

\bibitem{JZhu:14}
J.~Zhu, R.~Schober, and V.~K. Bhargava, ``{Secure transmission in multicell
  massive MIMO systems},'' {\em IEEE Trans. Wireless Commun.}, vol.~13,
  pp.~4766--4781, Sep. 2014.

\bibitem{JZhu:16}
J.~Zhu, R.~Schober, and V.~K. Bhargava, ``{Linear precoding of data and
  artificial noise in secure massive MIMO systems},'' {\em IEEE Trans. Wireless
  Commun.}, vol.~15, pp.~2245--2261, Mar. 2016.

\bibitem{ZJiang:15}
Z.~Jiang, A.~F. Molisch, G.~Caire, and Z.~Niu, ``{Achievable rates of FDD
  massive MIMO systems with spatial channel correlation},'' {\em IEEE Trans.
  Wireless Commun.}, vol.~14, pp.~2868--2882, May 2015.

\bibitem{MFrikel:00}
M.~Frikel, W.~Utschick, and J.~Nossek, ``{Blind noise and channel
  estimation},'' in {\em Proc. IEEE 10th Workshop on Statistical Signal and
  Array Processing}, 2000.

\bibitem{YChi:10}
Y.~Chi, Y.~Wu, and R.~Calderbank, ``{Regularized blind detection for MIMO
  communications},'' in {\em Proc. IEEE ISIT 2010}, Jul. 2010.

\bibitem{MBloch:11}
M.~Bloch and J.~Barros, {\em {Physical layer security: From information theory
  to security engineering}}.
\newblock Cambridge Universe Press, 2011.

\bibitem{CLing:14}
C.~Ling, L.~Luzzi, J.-C. Belfiore, and D.~Stehle, ``{Semantically secure
  lattice codes for the Gaussian wiretap channel},'' {\em IEEE Trans. on Inf.
  Theory}, vol.~60, pp.~6399--6416, Sep. 2014.

\bibitem{XLi:07}
X.~Li, J.~Hwu, and E.~P. Ratazzi, ``{Using antenna array redundancy and channel
  diversity for secure wireless transmissions},'' {\em J. Commun.}, vol.~2,
  pp.~24--32, May 2007.

\bibitem{YKozai:15}
Y.~Kozai and T.~Saba, ``{An artificial fast fading generation scheme for
  physical layer security of MIMO-OFDM systems},'' in {\em Proc. IEEE ICSPS
  2015}, Jan. 2015.

\bibitem{HMWang:15}
H.~M. Wang, T.~Zheng, and X.~G. Xia, ``{Secure MISO wiretap channels with
  multiantenna passive eavesdropper: Artificial noise vs. artificial fast
  fading},'' {\em IEEE Trans. Wireless Commun.}, vol.~14, pp.~94--106, Jan.
  2015.

\bibitem{CSong:18TVT}
C.~Song, ``{Achievable secrecy rate of artificial fast fading techniques and
  secret-key assisted design for MIMO wiretap channels with multi-antenna
  passive eavesdropper},'' {\em to appear in IEEE Trans. Veh. Technol.}

\bibitem{TMarzetta:99}
T.~L. Marzetta and B.~M. Hochwald, ``{Capacity of a mobile multiple-antenna
  communication link in Rayleigh flat fading},'' {\em IEEE Trans. Inf. Theory},
  vol.~45, pp.~139--157, Jan. 1999.

\bibitem{LZheng:02}
L.~Zheng and D.~N.~C. Tse, ``{Communication on the Grassmann manifold: A
  geometric approach to the non-coherent multiple-antenna channel},'' {\em IEEE
  Trans. Inf. Theory}, vol.~48, pp.~359--383, Feb. 2002.

\bibitem{SSamai:02}
S.~Shamai and T.~L. Marzetta, ``{Multiuser capacity in block fading with no
  channel state information},'' {\em IEEE Trans. Inf. Theory}, vol.~48,
  pp.~938--942, Apr. 2002.

\bibitem{HJi:16}
H.~Ji, Y.~Kim, J.~Lee, E.~Onggosanusi, Y.~Nam, J.~Zhang, B.~Lee, and B.~Shim,
  ``{Overview of full-dimension MIMO in LTE-Advanced Pro},'' {\em IEEE Commun.
  Mag.}, vol.~55, pp.~176--184, Feb. 2017.

\bibitem{AMukherjee:12}
A.~Mukherjee and A.~L. Swindlehurst, ``{Detecting passive eavesdroppers in the
  MIMO wiretap channel},'' in {\em Proc. IEEE Int. Conf. Acoust., Speech Signal
  Process. (ICASSP)}, pp.~2809--2812, Mar. 2012.

\bibitem{YChoi:15}
Y.~Choi and D.~Kim, ``{Performance analysis with and without torch node in
  secure communications},'' in {\em Proc. IEEE Int. Conf. Adv. Technol. Commun.
  (ATC)}, pp.~84--87, Oct. 2015.

\bibitem{DMaiwald:00}
D.~Maiwald and D.~Kraus, ``{Calculation of moments of complex Wishart and
  complex inverse Wishart distributed matrices},'' {\em IEE Proc. Radar Sonar
  Navig.}, vol.~147, pp.~162--168, Aug. 2000.

\bibitem{AKhisti:10a}
A.~Khisti and G.~W. Wornell, ``{Secure transmission with multiple antennas -
  PartII: The MIMOME wiretap channel},'' {\em IEEE Trans. Inf. Theory},
  vol.~56, pp.~5515 -- 5532, Nov. 2010.

\bibitem{WMBoothby:86}
W.~M. Boothby, {\em {An introduction to differential manifolds and Riemannian
  geometry}}.
\newblock 2nd ed. San Diego, CA: Academinc, 1986.

\bibitem{PMLee:97}
P.~M. Lee, {\em {Bayesian statistics: An introduction}}.
\newblock 2nd ed, Arnold/John Wiley, London/NewYork, 1997.

\bibitem{BHassibi:03}
B.~Hassibi and B.~M. Hochwald, ``{How much training is needed in
  multiple-antenna wireless links?},'' {\em IEEE Trans. Inf. Theory}, vol.~49,
  pp.~951--963, Apr. 2003.

\bibitem{HSi:15}
H.~Si, O.~O. Koyluoglu, and S.~Vishwanath, ``{Achieving secrecy without any
  instantaneous CSI: polar coding for fading wiretap channels},'' in {\em Proc.
  2015 IEEE International Symposium on Information Theory (ISIT)},
  pp.~2161--2165, 2015.

\bibitem{TAnderson:58}
T.~Anderson, {\em {An introduction to Multivariate Statistical Analysis}}.
\newblock New York: Wiley, 1958.

\end{thebibliography}

\end{document}